\documentclass[
 reprint, 
 amsmath, 
 amssymb, 
 aps, 
 balancelastpage,
 floatfix,
 superscriptaddress 
]{revtex4-2}

\newcommand{\adag}{\hat{a}^\dagger}
\newcommand{\ahat}{\hat{a}}

\newcommand{\aphat}{\hat{a}^\prime}
\newcommand{\sz}{\hat{\sigma}_z}
\newcommand{\Nf}{\hat{N}_f}
\newcommand{\phif}{\hat{\phi}_f}
\newcommand{\omr}{\omega_r}
\newcommand{\omd}{\omega_d}
\newcommand{\p}{\prime}
\newcommand{\nbar}{\Bar{n}}

\newcommand{\deleff}{\Delta_{\mathrm{eff}}}
\newcommand{\chieff}{\chi_{\mathrm{eff}}}
\newcommand{\modalpha}{|\alpha|^2}             
\newcommand{\argalpha}{\left(|\alpha|^2\right)} 
\newcommand{\Hhat}{\hat{H}}
\newcommand{\Ud}{\hat{U}_d}

\providecommand{\Eprint}{\href}
\usepackage{graphicx}
\usepackage{dcolumn}
\usepackage{bm}
\usepackage{lipsum}
\usepackage{xcolor}          
\usepackage{placeins}
\usepackage{physics}
\usepackage[caption=false]{subfig} 

\usepackage{tikz}
\usetikzlibrary{svg.path}
\definecolor{lime}{HTML}{A6CE39}
\newcommand{\orcid}[1]{\href{https://orcid.org}{\begin{tikzpicture}[overlay,remember picture]
\draw[coordinate,yshift=0.2ex] (0,0) coordinate (O);
\fill[lime] (O) circle (0.11cm);
\draw[white,line width=0.02cm] (O) circle (0.11cm);
\fill[white] (O) ++(-0.025cm,0.05cm) rectangle ++(0.015cm,-0.1cm);
\fill[white] (O) ++(0.015cm,0.05cm) circle (0.01cm);
\fill[white] (O) ++(0.015cm,0.025cm) rectangle ++(0.015cm,-0.075cm);
\end{tikzpicture}\hspace{8pt}}}

\usepackage[unicode=true, bookmarks=false, breaklinks=true, pdfborder={0 0 1}, backref=false, colorlinks=true]{hyperref}
\hypersetup{
     linkcolor=red, 
     urlcolor=blue,
     citecolor=blue,
     hypertexnames=true
}

\begin{document}

\preprint{APS/123-QED}

\title{Bosonic quantum control with a weakly coupled fluxonium qubit}

\author{Anaida~Ali}
\affiliation{Département de Physique and Institut Quantique, Université de Sherbrooke, Sherbrooke, Québec, J1K 2R1, Canada}

\author{Shantanu~R.~Jha}
\affiliation{Research Laboratory of Electronics, Massachusetts Institute of Technology, Cambridge, Massachusetts 02139, USA}
\affiliation{Department of Electrical Engineering and Computer Science, Massachusetts Institute of Technology, Cambridge, Massachusetts 02139, USA}

\author{Shoumik~D.~Chowdhury}
\affiliation{Research Laboratory of Electronics, Massachusetts Institute of Technology, Cambridge, Massachusetts 02139, USA}
\affiliation{Department of Electrical Engineering and Computer Science, Massachusetts Institute of Technology, Cambridge, Massachusetts 02139, USA}

\author{Lev-Arcady~Sellem}
\affiliation{Département de Physique and Institut Quantique, Université de Sherbrooke, Sherbrooke, Québec, J1K 2R1, Canada}

\author{Max~Hays}
\affiliation{Research Laboratory of Electronics, Massachusetts Institute of Technology, Cambridge, Massachusetts 02139, USA}

\author{William~D.~Oliver}
\affiliation{Research Laboratory of Electronics, Massachusetts Institute of Technology, Cambridge, Massachusetts 02139, USA}
\affiliation{Department of Electrical Engineering and Computer Science,
Massachusetts Institute of Technology, Cambridge, Massachusetts 02139, USA}
\affiliation{Department of Physics, Massachusetts Institute of Technology, Cambridge, Massachusetts 02139, USA}

\author{Baptiste~Royer}
\affiliation{Département de Physique and Institut Quantique, Université de Sherbrooke, Sherbrooke, Québec, J1K 2R1, Canada}

\begin{abstract}
Echoed Conditional Displacement $\left(\text{ECD}\right)$ gates constitute a fundamental building block for quantum control of harmonic oscillator modes. However, bit-flips of the auxiliary qubit remain a dominant error mechanism for this kind of bosonic control. In this work, we present a numerical case study of a bit-flip protected fluxonium operating as the control qubit and numerically implement ECD gates in a single-mode resonator-fluxonium device, demonstrating that fidelities exceeding $99.9\%$ are possible. We systematically investigate the resonator dynamics using a combination of semiclassical trajectories and master equation simulations, numerically revealing asymptotic saturation of the dispersive shift in the strongly driven regime of the resonator.
We develop an efficient technique to numerically simulate the strongly driven regime of the resonator using a semiclassical formulation that maps the full perturbation series in the dispersive expansion as order-by-order frequency shifts. This provides a compact polynomial description of the resonator which is intuitive and remains valid throughout the dispersive regime. Furthermore, we propose an improved ECD sequence that accounts for the effects of photon loss and spurious nonlinear terms on resonator trajectories.
\end{abstract}

\maketitle
\section{\label{sec:main level 1}INTRODUCTION}

Bosonic codes offer a hardware-efficient route towards fault-tolerance, encoding logical information in the continuous phase space of one or many quantum harmonic oscillators~\cite{Fluhmann_2019, Ofek_2016, Hu_2019, Campagne_Ibarcq_2020,deNeeve_2022, Liu_2026}. These codes can be realized in various systems such as electromagnetic modes of microwave resonators or motional modes of trapped ions.
However, such linear systems do not admit universal control by themselves;
the ability to harness the large Hilbert space of oscillators is only made possible by coupling them to nonlinear elements such as auxiliary qubits. The control problem in such a hybrid qubit-oscillator system can then be stated as: how can we control long-lived linear oscillators with qubits that typically have shorter lifetimes? Several hybrid qubit-oscillator algorithms have also been proposed~\cite{Liu_2026, Singh_2026, Martyn_2021, Crane_2026}, further increasing the need for fast and high-fidelity control methods. 

Minimal device footprint and favorable error hierarchies of high-quality microwave resonators establish superconducting circuits as an attractive testbed to explore bosonic control~\cite{Ni_2023, Sivak_2023, Lescanne_2020, Brock_2025}. In such a long-lived resonator-qubit circuit, coherences are often limited by the short lifetimes of auxiliary qubits~\cite{Putterman_2025, Royer2020, Royer2022}. Additionally, the storage resonator inherits spurious nonlinearities from the control qubit that can dephase the logical information.
Attempts to fulfill the wishlist of quantum control---finding strategies that preserve the error hierarchies of resonators while maintaining high speed of control---have resulted in the realization of several robust control schemes like the phase space Instruction Set Architecture $\left(\text{ISA}\right)$, the Fock ISA, SNAP control, and others~\cite{Ma_2021, Eickbusch_2022, Diringer_2024, Heeres_2017}. 

The Echoed Conditional Displacement (ECD) gate constitutes the phase space ISA, which achieves the aforementioned control wishlist by exploiting a weak-coupling-large-displacement architecture, where large phase-space displacements compensate for small values of qubit-resonator coupling strengths~\cite{Eickbusch_2022}. Weak coupling reduces qubit-induced nonlinearities in the resonator while large displacements effectively amplify the desired qubit-resonator interaction and maintain fast quantum control. Along with single-qubit rotations, ECD gates constitute a universal gate set, making them a valuable tool in bosonic quantum control. This is especially true for Gottesman-Kitaev-Preskill $\left(\text{GKP}\right)$ codestates~\cite{Gottesman_2001}, as their translationally symmetric structure in phase space makes ECD gates a natural gate choice. Typically, the fidelity of ECD gates is limited by the control qubit bit flips, while by contrast, the echo pulse in the ECD sequence intrinsically protects the gate from slow qubit dephasing. This suggests a direct pathway to improve the control: using control qubits that are engineered to protect against the qubit relaxation channel~\cite{Puri_2019, Ding_2025, Zheng_2025, Nie2026}. 

In this work, we present a numerical case study of ECD gates where the control qubit is a fluxonium operating at the half-flux sweet spot where coherence is maximized while retaining sufficiently long relaxation times. In the phase basis, fluxonium exhibits a double-well confinement near half-flux which suppresses qubit bit-flips from the excited to the ground state. This can lead to the fluxonium possessing substantially improved relaxation times in excess of a few milliseconds~\cite{Manucharyan_2009, Zhu_2013, Ding_2023}. 
We provide a comprehensive study of resonator and fluxonium dynamics in the large-displacement limit, where the system begins to deviate from its dispersive form. Furthermore, we outline strategies to improve the overall performance of ECD gates and derive an improved ECD pulse sequence that accounts for resonator photon loss and higher-order resonator nonlinearities. 

This paper is organized as follows: in Sec.~\ref{sec:main level 2}, we briefly review the theory underlying ECD gates. In Sec.~\ref{sec:main level 3}, we characterize the resonator-fluxonium device, and in Sec.~\ref{sec:main level 4}, we evaluate the performance of the ECD gate and discuss the dominant error mechanisms limiting ECD fidelities in the simulated device. In Sec.~\ref{sec:main level 5}, we present an improved ECD sequence starting from an effective Hamiltonian model that includes resonator photon loss and higher-order nonlinear terms which are not accounted for in the standard ECD protocol. Sec.~\ref{sec:main level 6} summarizes this work.
\begin{figure*}[tp]
    \centering
    \includegraphics[scale = .5]{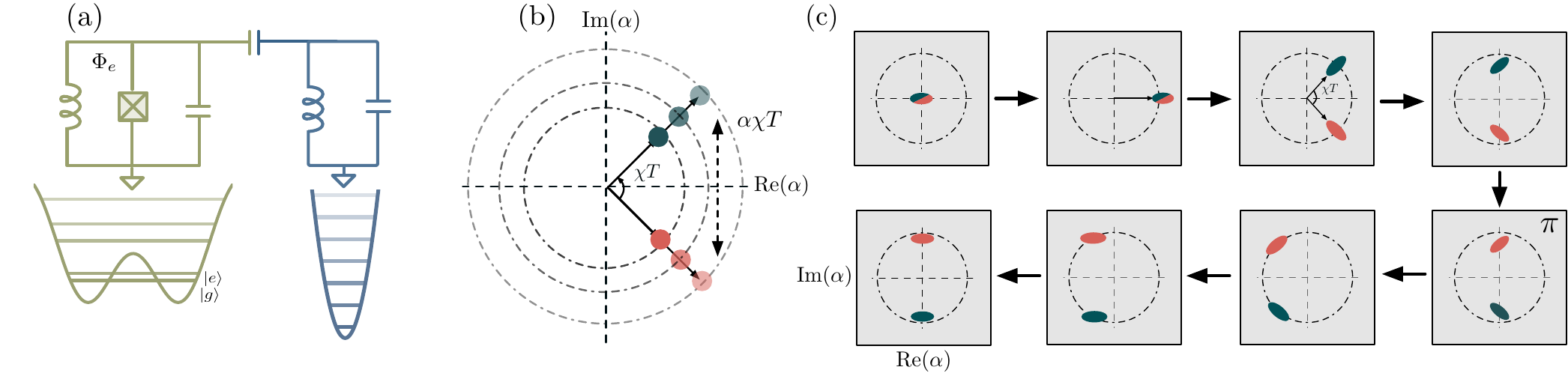}
        \caption{$\left(\text{a}\right)$ Circuit schematic of a lumped-element resonator (blue) coupled capacitively to a fluxonium (green). $\left(\text{b}\right)$ Under a controlled-rotation interaction, angular velocity remains fixed at $\chi$, while the separation $\alpha\chi T$ between the resonator states when the qubit is in $\ket{g}$ (blue) and $\ket{e}$ (red) increases with $|\alpha|$. $\left(\text{c}\right)$ Phase-space trajectory of an ECD gate with qubit in $\ket{+} = \frac{\ket{g} + \ket{e}}{\sqrt{2}}$ and resonator starting in squeezed vacuum.}
        \label{fig:circuit illustration}
\end{figure*}
\noindent
\section{\label{sec:main level 2}THEORY OF ECHOED CONDITIONAL DISPLACEMENT GATE}
We first consider an idealized circuit QED system consisting of a microwave resonator coupled to a control qubit. In the doubly rotating frame of the resonator and the qubit, the ideal dispersive Hamiltonian is expressed as
\begin{equation}
    \Hhat = \frac{\chi}{2}\adag\ahat\sz + \epsilon^* \ahat + \epsilon \adag,
    \label{dispersive_ecd_hamiltonian}
\end{equation}
where the resonator and qubit modes are represented by the annihilation operator $\ahat$ and Pauli operator $\sz = \ketbra{e}{e} - \ketbra{g}{g}$, respectively. The resonator is driven by a time-varying linear drive of strength $\epsilon$. A dispersive interaction of strength $\chi$ couples the resonator and qubit modes, enabling coherent control of the resonator via the qubit~\cite{Blais2004, Blais2021}. This suggests a timescale  $\pi/\chi$ for control, meaning that a large value of $\chi$ enables fast control. However, absent from the above equation is the Kerr interaction with strength $K\propto \chi^2$, which induces undesired evolution in phase space. Reducing $\chi$ at the expense of slow gate times does not help to improve fidelities either -- if $\chi$ is too low, the control speed becomes comparable to decoherence in the system~\cite{Heeres_2015, Leghtas2013}, leading to a trade-off between gate speed and control fidelity. ECD gates breaks this trade-off by leveraging large phase-space displacements~\cite{EickbuschThesis2024} to selectively amplify the desired interaction. 
This effect is best shown by going into the displaced frame of the resonator in which the linear drive term above vanishes, $\ahat \rightarrow \ahat + \alpha$, where $\alpha$ is the displacement of the resonator relative to the origin of phase space $\left(\text{see Appendix~\ref{appendix:level2}}\right)$. In this displaced frame, Eq.~\eqref{dispersive_ecd_hamiltonian} becomes
\begin{equation}
    \hat{H}_{\mathrm{displaced}} = \frac{\chi}{2}\adag\ahat\sz + \frac{\chi}{2}|\alpha|^2\sz + \frac{\chi}{2}\left(\adag\alpha + \ahat\alpha^*\right)\sz.
    \label{displaced_ecd_Hamiltonian}
\end{equation}
The third term above is the generator of the desired resonator-qubit entangling operation -- conditional displacement with an enhanced interaction rate $\chi|\alpha|$. As illustrated in Fig.~\ref{fig:circuit illustration}(b), this leads to a qubit-controlled displacement of the resonator with an effective strength $g_{\mathrm{CD}} = \chi|\alpha|$ with $|\alpha| \gg 1$, amplifying the weak dispersive strength $\chi$ by a large phase-space displacement $|\alpha|$~\cite{Eickbusch_2022}. However, the first two terms prevent the realization of an ideal conditional displacement operation due to the always-on dispersive coupling and qubit Stark shift, respectively. 
These two terms are cancelled by applying an echo on the qubit and the resonator, which can be expressed as $\sz \rightarrow -\sz$ and $\alpha \rightarrow -\alpha$, respectively, resulting in a purely conditional displacement interaction. The full ECD sequence is illustrated in Fig.~\ref{fig:circuit illustration}(c), and the unitary yielding the ECD gate has the form
\begin{equation}
    \text{ECD}\left(\beta\right) = \hat{D}\left(\frac{\beta}{2}\right)\ketbra{e}{g} + \hat{D}\left(-\frac{\beta}{2}\right)\ketbra{g}{e},
    \label{ecd_unitary}
\end{equation}
where $\hat{D}\left(\alpha\right) = e^{\alpha \adag - \alpha^*\ahat}$ is the phase-space displacement operator. For a total gate time $T$, the desired conditional displacement $\beta \approx \alpha\chi T$ ($\chi T < 2\pi$) can be achieved with a resonator drive of the form ~\cite{Eickbusch_2022, EickbuschThesis2024}
\begin{equation}
    \begin{aligned}
    \epsilon\left(t\right) &= {}\alpha\Bigg[\delta\left(t\right) - 2\cos{\left(\frac{\chi T}{4}\right)}\delta\left(t - \frac{T}{2}\right) +\\&{} \delta\left(t-T\right)\cos{\left(\frac{\chi T}{2}\right)}\Bigg],
    \label{ecd_pulse_sequence}
    \end{aligned}
\end{equation}
with the Dirac delta functions in practice approximated using, for example, short Gaussian pulses. Note that the ECD sequence considered here flips the qubit state, and a pure conditional displacement is easily recovered by adding a final qubit $\pi$ pulse.
\section{\label{sec:main level 3}CHARACTERIZING THE RESONATOR-FLUXONIUM SYSTEM}
We now turn to our system of interest, a resonator capacitively coupled to a fluxonium qubit, as illustrated in Fig.~\ref{fig:circuit illustration}(a). The Hamiltonian of the system can be written as~\cite{Zhu_2013, Blais2021}, 
\begin{equation}
    \begin{aligned}
    \hat{H}_{\mathrm{rf}} ={}& \omr^{\p}\hat{a}^{\prime \dagger}\hat{a}^{\prime} + 4E_C\Nf^2 - E_J\cos{\left(\phif - \frac{2\pi \Phi_e}{\Phi_0}\right)} + \\
    &\frac{E_L}{2}\phif^2 + ig\Nf\left(\hat{a}^{\prime \dagger} - \hat{a}^{\prime}\right) + 2\epsilon\cos{\omd t}\left(\hat{a}^{\prime \dagger} + \hat{a}^{\prime}\right),
    \label{rf_hamiltonian}
    \end{aligned}
\end{equation}
where $\omr^\p$ is the bare frequency of the resonator associated with the bare annihilation operator $\aphat$, $\omd$ is the center frequency of the slowly varying driving tone $\epsilon$, and the resonator is coupled to the fluxonium qubit with an interaction strength $g$. The parameters $E_C$, $E_L$, and $E_J$ are the charging, inductive, and Josephson energies of the fluxonium qubit, respectively. The charge and phase operators of the fluxonium are given by $\Nf$ and $\phif$, respectively, and $\Phi_e$ is the external flux threading the loop formed by the Josephson junction and the inductor, with $\Phi_0 = h/2e$ being the superconducting flux quantum.

We choose our operating point for the external flux at half-flux, $\Phi_e = \Phi_0/2$,
where the fluxonium features a double-well potential occupied by nearly degenerate energy states. The degeneracy is broken by tunneling between the two wells, resulting in typical qubit frequencies near half-flux on the order of the thermal energy scales. Near half-flux, ground and excited state wavefunctions of the fluxonium are localized in different potential wells, and the spatial separation of wavefunctions suppresses bit-flip transitions between the qubit states, resulting in fluxonium possessing $T_1$ lifetimes in the order of milliseconds close to half-flux ~\cite{Somoroff2023, Lin_2018, Zhang_2021}. For heavy-fluxonium qubits, like the device studied in this work, $T_1$ typically drops at half-flux due to thermal excitations~\cite{Zhang_2021, siegele_2025, Gyenis_2021}. Nevertheless, due to the half-flux being a flux sweet spot, this operating point benefits from first-order suppression against flux dephasing while still retaining reasonable $T_1$ values close to a millisecond~\cite{Nguyen2019, Somoroff2023, Jha_2026}. Furthermore, the rich, flux-tunable spectrum of fluxonium removes the strict relationship between dispersive shift $\chi$ and resonator self-Kerr, unlike transmon qubits~\cite{Zhu_2013, Blais2021, Nie2026}. 

In the dispersive regime, at low photon number and under the rotating-wave approximation, Eq.~\eqref{rf_hamiltonian} takes the form~\cite{Minev_2021, Eickbusch_2022}
\begin{equation}
    \begin{aligned}
        &\hat{H}_{\mathrm{rf}} \approx \omega_r \hat{a}^{\dagger}\hat{a} \ \ \left\} \text{      \footnotesize linear}\right. \\ 
        &\left. \begin{aligned}
            &+ \sum_m \omega_m\ketbra{m}{m} - K_c\hat{a}^{\dagger^2}\hat{a}^2 - \\
            &\sum_l \left(\chi_m\hat{a}^{\dagger}\hat{a} + \chi^{\prime}_m\hat{a}^{\dagger^2}\hat{a}^2\right) \ketbra{m}{m} 
        \end{aligned} \right\} \ \ \text{\footnotesize nonlinear} \\ 
        &+ 2\epsilon\left(t\right)\cos{\omega_d t}\hat{N}_r \ \ \left\} \ \ \text{\footnotesize drive}\right..
    \end{aligned}
    \label{dressed_rf_hamiltonian_dispersive}
\end{equation}

\noindent
Here $\ahat$ denotes the annihilation operator of the resonator-like mode, and $\ket{m}$ enumerates fluxonium-like eigenmodes with the ground state labeled by $m = 0$. We denote the dressed resonator frequency by $\omr$ and the fluxonium energies by $\omega_m$. $K_c$ is the resonator self-Kerr, and $\chi$ and $\chi^{\prime}$ are the dispersive shifts arising from fourth- and sixth-order perturbative expansions, respectively. $\hat{N}_r$ is the dressed resonator charge operator.  In this work, unless stated otherwise, we simulate the device labeled Device 1 with the following parameters:  $E_J/2\pi = 2.756$ GHz, $E_C/2\pi = 856.8$ MHz, $E_L/2\pi = 301.8$ MHz, yielding a qubit frequency at half flux of $\omega_{\mathrm{g-e}}/2\pi = 233$ MHz. 
The resonator is at a frequency of $\omega_r/2\pi = 4.146$ GHz with a loss rate $\kappa/2\pi = 6.37$ kHz and $g/2\pi = 6.329$ MHz, yielding dispersive shifts $\chi/2\pi = 32.77$ kHz and $\chi^{\prime}/2\pi = 3.5$ Hz, and a self-Kerr nonlinearity of $K_c/2\pi = 3.5$ Hz at half flux  $\left(\text{see Table \ref{Table_device_parameters} in Appendix~\ref{appendix:level3}}\right)$. 
\begin{figure*}[tp]
    \centering
    \includegraphics[width = .9\linewidth]{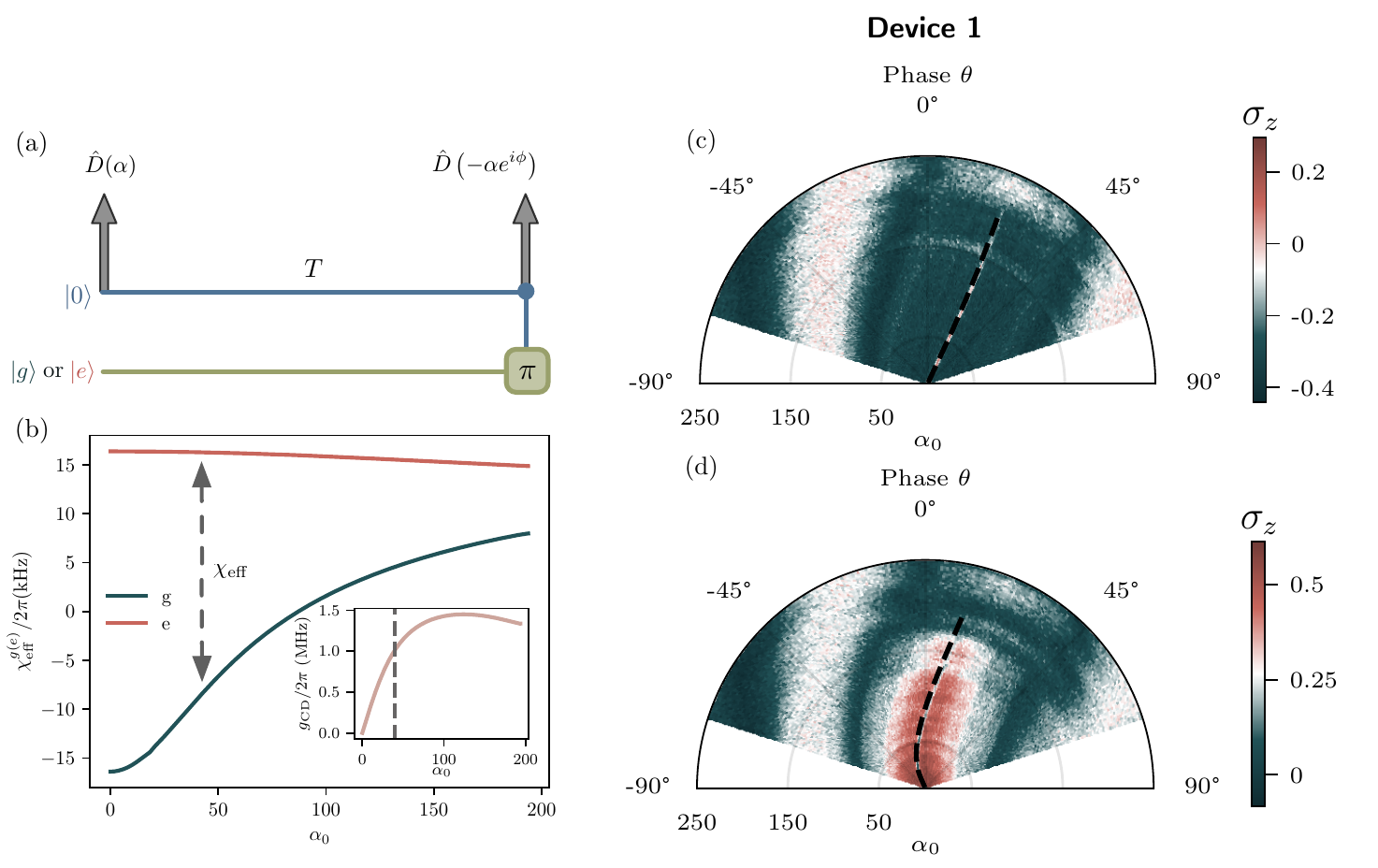}
    \captionsetup{width=1\linewidth}
    \caption{$\left(\text{a}\right)$ \textit{out-and-back} sequence to measure the phase accumulated by the resonator with the qubit initialized in $\ket{g}$ or $\ket{e}$. $\left(\text{b}\right)$ Photon number-dependent resonator dispersion obtained from \textit{out-and-back} where the net frequency pull on the resonator by the qubit $\chieff$ is given by the separation between the two curves. The average of the two curves is the effective detuning $\deleff$ felt by the resonator. The inset shows the ECD interaction rate $g_{\mathrm{CD}} = \chieff|\alpha_0|$ as a function of displacement. Dashed gray lines indicate the regime after which the system Hamiltonian is described better by the model in Eq.~\eqref{full_dispersive_hamiltonian}. $\left(\text{c, d}\right)$ Numerically obtained dispersion (black dashed lines) from \textit{out-and-back} performed on Device 1, overlaid on top of measured data obtained in the experiments of Ref.~\cite{Jha_2026}.  Fluxonium is initialized in $\ket{g}(\text{top})$ and $\ket{e}(\text{bottom})$ respectively $(\text{see Appendix~\ref{appendix:level3}})$. The loss of contrast observed far from the return phases is likely due to drive-induced state transitions from the large excess of photons occupying the storage resonator, which in turn causes assignment errors \cite{Dumas_2024}.}
    \label{fig:out and back}
\end{figure*}
\noindent
The correct choice of drive frequency is crucial for achieving the desired dynamics expected from ECD, as the center frequency $\omd$ of the drive has to be set at a detuning of $\left(\chi_g + \chi_e\right)/2$ from the dressed resonator frequency, i.e., $\omr - \omd = \left(\chi_g + \chi_e\right)/2$~\cite{Eickbusch_2022}. Therefore, careful estimation and calibration of the Hamiltonian parameters $\chi$ and Kerr are paramount to ECD. These parameters can be computed via numerical diagonalization of Eq.~\eqref{rf_hamiltonian} using the relation $\chi_{\mathrm{g\left(\text{e}\right)}} =  E_{\mathrm{{g\left(\text{e}\right)}, 0}} - E_{\mathrm{{g\left(\text{e}\right)}, 1}}$ and $2K_c = 2E_{\mathrm{g, 1}} - E_{\mathrm{g, 2}} - E_{\mathrm{g, 0}}$, where $E_{\mathrm{q, n}}$ denotes the eigenenergies with the first and second indices representing the fluxonium and resonator state labels, respectively $\left(\text{see Appendix \ref{appendix:level1}}\right)$.

At high resonator photon numbers, however, higher-order nonlinearities become important, and the usual Hamiltonian in Eq.~\eqref{dressed_rf_hamiltonian_dispersive} does not describe the dynamics of the system well. In a doubly rotating frame of the resonator and the fluxonium qubit, the system given by Eq.~\eqref{rf_hamiltonian}  is better described by 
\begin{equation}
\begin{aligned}
        \Hhat_{\mathrm{rf}} &= \Delta \adag\ahat - \sum_{n = 2}{K^{\left(2n\right)} \hat{a}^{\dagger n}\hat{a}^{n}} \\
        &- \sum_{n = 1, m = 0}\chi^{\left(2n + 2, m\right)}\hat{a}^{\dagger n}\hat{a}^{n}\ketbra{m}{m},
    \label{full_dispersive_hamiltonian}
\end{aligned}
\end{equation}
where $K^{\left(2n\right)}$ is the $2n$-th order resonator kerr-like term and $\chi^{\left(2n + 2, m\right)}$ is the $2\left(n + 1\right)$-th order dispersive frequency pull between the resonator and $m$-th fluxonium level. Under a displacement $\ahat \rightarrow \ahat + \alpha$, these terms give rise to several lower-order terms. The $2n$-th ordered Kerr-like interaction results in a detuning amplified by a factor of $|\alpha|^{2\left(n-1\right)}$ 
\begin{equation}
K^{(2n)}\xrightarrow[]{\mathrm{detuning \hspace{.1 cm}of}} \left(-1\right)^nn|\alpha|^{2\left(n-1\right)}|K^{(2n)}|, 
\label{phasespace_enhancement}
\end{equation}
caused by the dressing of higher-order nonlinearities under the resonator drive, with a similar amplification for the dispersive terms. This drive-induced strengthening of higher-order terms necessitates a careful treatment accounting for the full dispersive model in the strongly driven regime of the resonator (see detailed treatment in Appendix~\ref{appendix:level2}). Confining Eq.~\eqref{full_dispersive_hamiltonian} into the dressed qubit subspace using the relation $\sz = \sum_{n}{\ketbra{e, n}{e, n} - \ketbra{g, n}{g, n}}$ where $\ket{q, n}$, $q = g, e$ are the dressed eigenstates identified with fluxonium and resonator indices $q$ and $n$ respectively (see Appendix~\ref{appendix:level1}). In the limit $\langle\adag\ahat\rangle\gg 1$, we model the system in Eq.~\eqref{full_dispersive_hamiltonian} in its displaced frame with the Hamiltonian
\begin{equation}
    \hat{H}_{\mathrm{rf, displaced}} \approx \left[\deleff \left( |\alpha|^2 \right) - \frac{ \chi_{\mathrm{eff}}\left( |\alpha|^2\right)}{2}\sz\right]\adag\ahat.
    \label{effective_displaced_hamiltonian}
\end{equation}
Here, $\Delta_{\mathrm{eff}}\left( |\alpha|^2\right) \text{ and } \chi_{\mathrm{eff}}\left( |\alpha|^2\right)$ are functions of $|\alpha|^2$ that capture the photon-number dependence of resonator frequencies to all orders in perturbation, in principle, as long as the system remains far from multiphoton resonances within the computational manifold~\cite{Shillito_2022, Dumas_2024}. The phase-space trajectory of the non-linear resonator described by Eq.~\eqref{effective_displaced_hamiltonian} is obtained by solving
\begin{equation}
     \partial_t\alpha = -i\Bigg[\left(\deleff\left(\modalpha\right) - \chieff\left(\modalpha\right)\frac{\langle\sz\rangle}{2}\right)\alpha + \epsilon\left(t\right)\Bigg] .
     \label{resonator_trajectory_maintext}
\end{equation}
While these higher-order nonlinearities can in principle be extracted from numerical diagonalization of the system Hamiltonian described in Eq.~\eqref{dressed_rf_hamiltonian_dispersive}, the system size makes it difficult for three reasons. Firstly, highly-excited states of the resonator need to be considered to extract higher-order perturbative terms, demanding a large Hilbert space to numerically describe the system, which requires heavy computational resources~\cite{Shillito_2022}. Secondly, it is unclear to what order the Hamiltonian should be truncated to obtain a good model for the system. Third, the expressions for higher-order terms beyond sixth order become lengthy and cumbersome to calculate analytically, limiting their usefulness~\cite{Xiao_2022, Fors2024}. To overcome these difficulties, we find it more convenient to numerically reproduce the experimental \textit{out-and-back} protocol to extract Hamiltonian parameters~\cite{Eickbusch_2022}.
Specifically, we numerically evolve the full resonator-fluxonium Hamiltonian of Eq.~\eqref{rf_hamiltonian} in a displaced frame and set $\deleff,\chieff$ such that solving the ODE given by Eq.~\eqref{resonator_trajectory_maintext} yields similar results for $\langle \hat a(t)\rangle$ (see Fig.~\ref{fig:semiclassical trajectories} in Appendix~\ref{appendix:level2}). This circumvents the aforementioned difficulties.

The \textit{out-and-back} sequence illustrated in Fig.~\ref{fig:out and back}(a) is described as follows. The qubit is initially prepared in the ground ($g$) or excited ($e$) state, and then a first pulse $\hat{D}\left(\alpha\right)$ displaces the resonator vacuum $\ket{0}\ket{g\left(e\right)}$ to a coherent state $\ket{\alpha_{g\left(e\right)} }\ket{g\left(e\right)}$. During the wait time $T$, it accumulates a phase $\phi_{g\left(e\right)}$ given by the effective frequency of the resonator, resulting in a state $\ket{\alpha_{g\left(e\right)} e^{i\phi_{g\left(e\right)}}}\ket{g\left(e\right)}$. A second pulse $\hat{D}\left(-\alpha e^{-i\theta}\right)$ attempts to bring the resonator back to vacuum, and the phase that brings the resonator back closest to vacuum measures the effective resonator dispersion. The dispersion obtained numerically from \textit{out-and-back} is highly nonlinear as shown in Fig.~\ref{fig:out and back}(b), indicating that higher-order nonlinearities undergo phase-space amplification and are activated at large displacements $\left(\text{see Appendix \ref{appendix:level2}}\right)$.  At low photon numbers ($|\alpha_0| < 30$), fitting the numerically obtained dispersion from \textit{out-and-back} to $n = 2$ in the model given by Eq.~\eqref{phasespace_enhancement} retrieves the quartic nonlinearities $\chi, K, \text{and } \chi^\p$, showing excellent agreement with numerical diagonalization. To validate our theory, we numerically simulated the devices used in the experiments in Ref.~\cite{Jha_2026} and our numerical results closely match the measured experimental data without any adjustable parameters, as shown in Fig.~\ref{fig:out and back}(c, d). 

The nonlinear resonator dispersion predicted by our numerical simulations in Fig.~\ref{fig:out and back}(b) and experimentally observed in Fig.~\ref{fig:out and back}(c, d) gives rise to several interesting observations. While Eq.~\eqref{displaced_ecd_Hamiltonian} predicts a linear increase in the interaction strength $g_{\mathrm{CD}}$ with $\alpha$, we observe a departure from this behavior as shown in the inset of Fig.~\ref{fig:out and back}(b). Here, $g_{\mathrm{CD}}$ increases with $|\alpha|$ up to a maximum value beyond which further displacement causes the interaction strength to degrade -- a behavior that cannot be captured by lower-order perturbative expansions like fourth-order dispersive treatments.
The origin of this behavior lies in the cosine potential of fluxonium -- the resonator term $K^{(2n)}$ arises from the $2n$-th order term ${\hat{\phi}}^{2n}$ of the cosine series which has a relative sign of $(-1)^n$ in the expansion, with a similar relation for $\chi^{\left(2n + 2, m\right)}$. The competing interplay between amplified terms of different orders, as expressed in Eq.~\eqref{phasespace_enhancement} results in the saturation of the dispersive shift, as shown in Fig.~\ref{fig:out and back}(b). This leads to the non-monotonic scaling of the ECD rate $g_{\mathrm{CD}} = \chieff\left(|\alpha|^2\right)|\alpha|$ with $|\alpha|$. Crucially, while perturbation terms beyond fourth order are nominally weak and can be safely neglected under standard operating conditions, they become important at large phase-space displacements. Merely extending the analysis to finite-order approximations—such as eighth- or tenth-order perturbation theory remains insufficient to explain this behavior. Our framework resolves this challenge through a semiclassical treatment of the resonator where we map the effects of the full perturbation series as order-by-order frequency shifts, which provides a compact polynomial description of the resonator dispersion. Our formulation provides a concise and intuitive description of the resonator and is valid throughout the dispersive regime where qubit transitions are suppressed.
\begin{figure}[t]
    \centering
    \includegraphics[width = 1\linewidth]{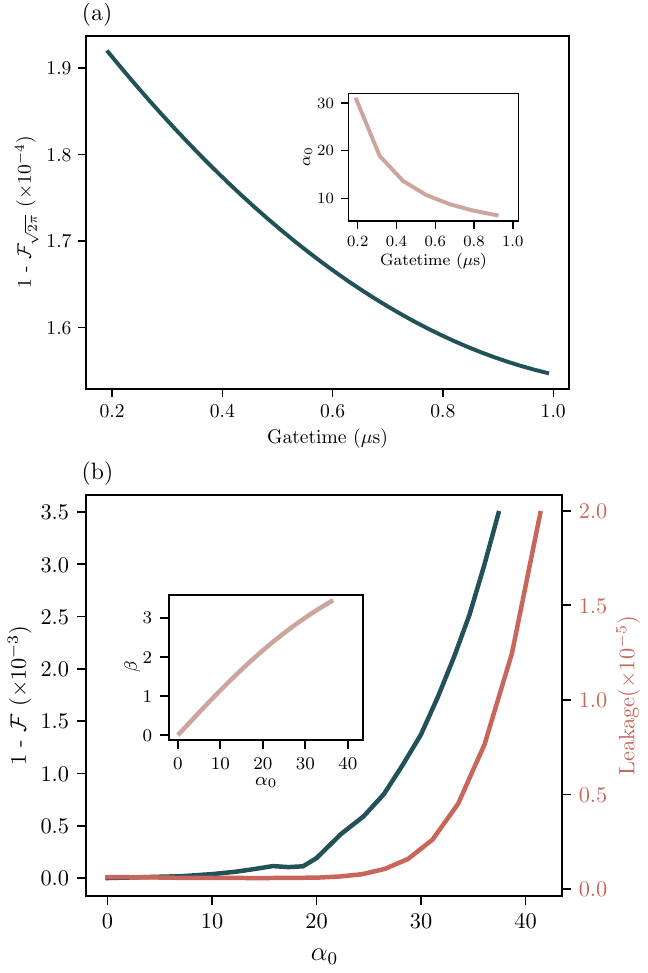}
    \captionsetup{width=1\linewidth}
    \caption{$\left(\mathrm{a}\right)$ Vacuum subspace fidelity as a function of gate time evaluated for a target displacement of $\beta = \sqrt{2\pi}$. Intermediate resonator displacement \(\alpha _{0}\) achieved for each gate time is indicated in the inset. $\left(\mathrm{b}\right)$ For a fixed gate time $T = 240$ ns, intermediate displacement $\alpha_0$ is varied resulting in different values of the conditional displacement $\beta$. The resulting fidelities (green) and fluxonium leakage (red) are plotted. Inset shows conditional displacements obtained for different values of $\alpha_0$.}
    \label{fig:vacuum fidelities}
\end{figure}
We verify the generality of our theoretical formulation by comparing the numerical results against experimental data obtained from different devices. Our theory shows excellent agreement and consistency across all the devices we tested (see Fig.~\ref{fig:out_back_other_devices}(a)-(d) in  Appendix~\ref{appendix:level2}). Remarkably, our framework captures experimentally observed features while restricting the resonator to just four Fock levels, which enables efficient time-domain simulations. Furthermore, this approach enables the exact simulation of high-photon-number regimes without the need for approximations or computational accelerators~\cite{Shillito_2022, Xiao_2022}.
\noindent
\section{\label{sec:main level 4}ECD WITH A FLUXONIUM CONTROL QUBIT}
Having numerically calibrated the effective dispersive parameters, we now simulate the full evolution of the ECD gate. We model the resonator drive $\epsilon$ in Eq.~\eqref{ecd_unitary} as a set of Gaussian pulses and retain the full model in Eq.~\eqref{rf_hamiltonian} without any approximations. We use perfect and instantaneous qubit $\pi$ echoes by applying an $X$ gate to the fluxonium, motivated by the existing single-qubit gate schemes obtaining high fidelities $>$99.999\%~\cite{Rower2024, Bao2022}. In Appendix~\ref{appendix:level6}, we present results for the $X$ gate implemented using $\pi$ echo pulses with fidelities exceeding $99.99\%$. We first aim to understand what errors fundamentally limit the performance of ECD in an ideal resonator-fluxonium device, so we ignore any dissipation in the system and solve the Schrödinger equation for the Hamiltonian in Eq.~\eqref{rf_hamiltonian}. We then analyze the implementation of ECD with a lossy resonator in detail in Sec.~\ref{sec:main level 5}.

We compute the subspace unitary fidelity of the ECD gate in the base case of the vacuum subspace described by the manifold spanned by $\{\ket{0}\}\otimes\{\ket{g}, \ket{e}\}$~\cite{Pedersen_2007, Zheng_2025}. Given that ECD gates form an important toolkit for GKP state manipulation, we determine the shortest gate duration to achieve a conditional displacement $\beta = \sqrt{2\pi}$ while maintaining fidelities close to 99.99\%.  This is motivated by the fact that the maximum displacement length involved in logical operations and GKP stabilization routines is $\sqrt{2\pi}$ ~\cite{Campagne_Ibarcq_2020, Royer2020, deNeeve_2022, Sivak_2023}. The resulting gate time is obtained by solving the time-dependent Schrödinger equation, starting from the resonator in the vacuum state. In Fig.~\ref{fig:vacuum fidelities}(a), we initialize the resonator in vacuum and fix our target displacement $\beta = \sqrt{2\pi}$. We then numerically compute the unitary fidelity as a function of gate time in the vacuum subspace, achieving fidelities exceeding $99.99\%$ in under 600 ns. 

Subsequently, we study ECD fidelity as a function of the phase space displacement of the resonator. In Fig.~\ref{fig:vacuum fidelities}(b), we fix the gate duration to 240 ns and vary the intermediate displacement $\alpha_0$, yielding different values of conditional displacement $\beta$ and recovered fidelities ranging from $99.9 - 99.99 \%$. We numerically evaluate the post-gate leakage population outside the computational manifold of the fluxonium qubit defined as $1 - \left(P_{g} + P_e\right)$, where $P_{g\left(e\right)}$ is the population in the ground $\left(\text{excited}\right)$ manifold of fluxonium. As shown in the inset figure of Fig.~\ref{fig:vacuum fidelities}(b), we find the leakage to be small. Rather, we attribute the decrease in fidelity to amplified resonator nonlinearities that distort the final coherent states (see Fig.~\ref{fig:overlap degradation} in Appendix~\ref{fidelities in vacuum subspace}). We remark that this happens at photon numbers well below so-called critical photon numbers~\cite{Schusterthesis2007, Blais2004} and below the maximum conditional displacement coupling strength computed in the previous section $\left(\text{inset Fig.~\ref{fig:out and back}}\left(\text{b}\right)\right)$.

\section{\label{sec:main level 5}ACCOUNTING FOR NONLINEARITIES AND PHOTON LOSS}
\begin{figure}[t]
    \centering
    \includegraphics[width=1\linewidth]{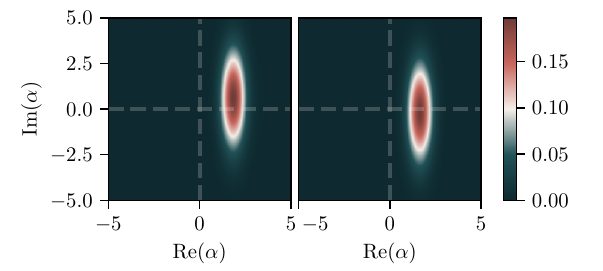}
    \caption{Wigner functions of the states generated by the standard ECD $\left(\text{left}\right)$ and improved ECD $\left(\text{right}\right)$ sequences. The improved ECD sequence cancels the unconditional displacement along the anti-squeezed quadrature. The simulated device has parameters of Device 1 and a $\kappa/2\pi = 160$ kHz. The resonator starts with 11 dB of squeezing.}
    \label{fig:wigner comparison}
\end{figure}
The standard ECD sequence is derived within fourth-order perturbation theory and also ignores (ideally negligible) resonator loss~\cite{Eickbusch_2022}. This regime is characterized by the linear scaling of the conditional displacement rate $g_{\mathrm{CD}}$ with phase-space displacement $|\alpha|$ as indicated by the region to the left of dashed grey lines in the inset of Fig.~\ref{fig:out and back}(b). However, the dynamics of the system described by Eq.~\eqref{rf_hamiltonian} will be slightly different from the ideal model described by Eq.~\ref{dressed_rf_hamiltonian_dispersive} due to the effect of higher-order nonlinear terms and photon loss on the resonator state. 
Moreover, higher-order terms $\left(\text{as illustrated in Fig.~\ref{fig:out and back}(b)}\right)$ and photon loss become stronger at larger photon numbers. This has two-fold effects -- first, activation of higher-order terms causes quantum state deformation~\cite{Leghtas2013, Heeres_2015, Kirchmair_2013}. Second, these unaccounted terms cause a deviation of the resonator phase-space trajectories from those expected from lower-order perturbation theories. Previous experiments accounted for these terms through numerical optimization of pulse parameters $\left(\text{see supplemental of Ref.~\cite{Eickbusch_2022}}\right)$.
If the nonlinearities are sufficiently weak such that they only cause a deviation of phase-space trajectories while preserving the shape of the state, we can still accurately model the phase-space dynamics of the resonator semiclassically by tracking the higher-order terms (see Fig.~\ref{fig:semiclassical trajectories} and Appendix~\ref{appendix:level2}). In this section, we propose a modified ECD sequence that is derived from a model that includes higher-order terms and resonator photon loss. Specifically, by modelling the actual dynamics of the system observed experimentally, we want to find the form of the pulse sequence that would realize a purely conditional displacement interaction. 

We consider a general drive of the form 
\begin{equation}
 \epsilon\left(t\right) = A_1\delta\left(t\right) + A_2\delta\left(t - \frac{T}{2}\right) + A_3\delta\left(t - T\right),
 \label{ecd_generic_pulse_sequence}
\end{equation}
\noindent
where $A_j$ are complex parameters to be determined. The effect of the qubit echo can be incorporated by dividing the middle pulse into two and inserting a $\pi$ pulse in the middle. To generate a conditional displacement $|\beta|e^{i\theta_{\beta}}$ with an intermediate displacement of $|\alpha_0|$, we must target drive amplitudes $A_j$ given by $\left(\text{see Appendix \ref{appendix:level4}} \text{ for detailed derivation}\right)$

\begin{subequations}
\label{improved_ecd_amplitudes}
\begin{align}
    A_1 &= |\alpha_0| e^{i \left(\theta_{\beta} + 2\deleff T\right)}, \\
    A_2 &= -|\alpha_0| \Bigg [\cos{\left(\frac{\chieff T}{4}\right)}  e^{-\frac{\kappa T}{4}} + \nonumber \\ 
    &\quad \sqrt{1 - \sin^2\left({\frac{\chieff T}{4}}\right) e^{\frac{-\kappa T}{2}}}\Bigg] e^{i \left(\theta_{\beta} + \frac{3}{2}\deleff T\right)}, \\ 
    A_3 &= -\left\{A_1 e^{-i\left(\deleff  - i\frac{\kappa}{2}\right)T}\right. \nonumber \\
    &\quad \left.+ A_2 e^{-i(\deleff - i\frac{\kappa}{2})\frac{T}{2}}\cos{\left(\frac{\chieff T}{4}\right)} \right\}.
\end{align}
\end{subequations}
The conditional displacement $\beta$ is given by
\begin{equation}
    \beta = 2A_2 e^{-i\left(\deleff - i\frac{\kappa}{2}\right)\frac{T}{2}}\sin{\left(\frac{\chieff T}{4}\right)}e^{-i\deleff T} .
    \label{S3E3}
\end{equation}
To evaluate the performance of our modified ECD, we first check its action on a squeezed state. In Fig.~\ref{fig:wigner comparison}, we compare the result of the standard ECD against our improved ECD sequence. The standard ECD gate suffers from residual unconditional displacement along the anti-squeezed axis that the sequence fails to cancel, and in contrast, the improved ECD gate generates only the target conditional displacement, whose value agrees closely with the theoretical prediction. 
Although this protocol does not correct for intrinsic nonlinear deformations of the resonator state, it successfully counteracts unwanted phase-space displacements and state rotations induced by loss and nonlinear terms (see Fig.~\ref{fig:average displacement} in Appendix~\ref{appendix:level4}). This becomes important when concatenating multiple ECD operations to synthesize a target unitary because any discrepancy between the expected and actual displacement accumulates over the rounds, potentially resulting in logical errors. On the other hand, our sequence ensures that these cascading errors are minimized during gate compilation, helping to simplify downstream calibration routines in experiments.
\begin{figure}[h!]
    \centering
    \includegraphics[width=1\linewidth]{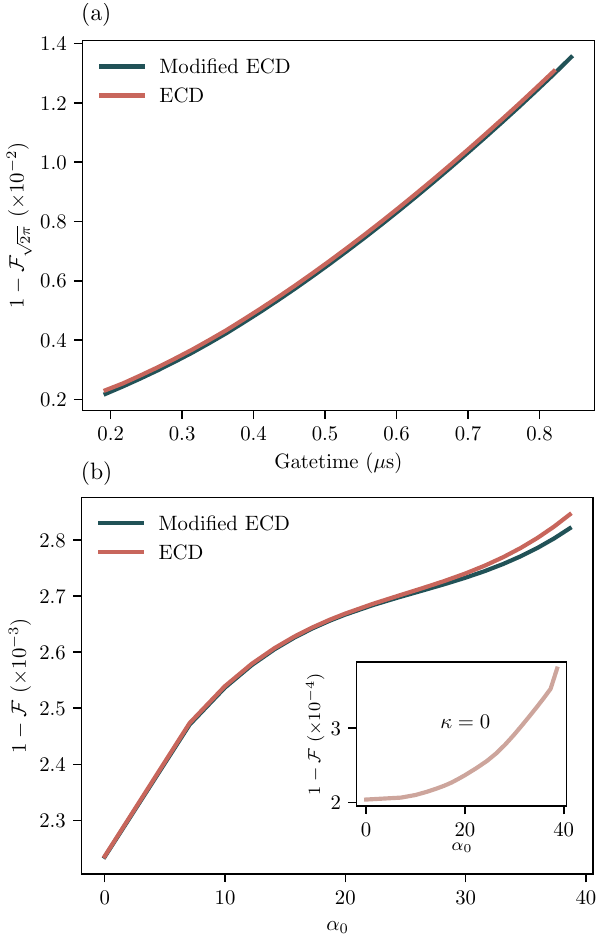}
    \caption{GKP subspace logical fidelities computed for $\left(\text{a}\right)$ $\beta = \sqrt{2\pi}$ as a function of gate time $\left(\text{b}\right)$ Fixing gate time $T = $ 240 ns, the applied displacement $\alpha_0$ is varied, resulting in different values of conditional displacement $\beta$, and fidelities are computed numerically using a bosonic subsystem decomposition ~\cite{Hopfmueller_2024}. Inset shows numerically obtained fidelities in the case of a lossless resonator.}
    \label{fig:modified ecd sequence}
\end{figure}
We compute the average ECD fidelities in the fluxonium-GKP product states described by $\{\ket{0_{\Delta}}, \ket{1_{\Delta}}\} \otimes \{\ket{g}, \ket{e}\}$, where $\ket{\mu_{\Delta}}$ corresponds to a GKP logical state $\mu$ with an average photon number $\nbar \approx \frac{1}{2}\left(\Delta^{-2} - 1\right)$ \cite{Royer2020}. The choice of subspace is motivated by the central role played by ECD gates in GKP stabilization -- consequently, ECD gate fidelities have a direct influence on logical lifetimes of GKP codestates~\cite{Sivak_2023, deNeeve_2022}. 

We perform time-dependent master equation simulations with product states of the resonator starting in the GKP code basis and the fluxonium in the computational basis. After partial tracing over the fluxonium, we numerically obtain the logical fidelities of the resulting resonator states using a subsystem decomposition of the bosonic Hilbert space~\cite{Hopfmueller_2024, Boudreault_2026}. While we evaluated the physical fidelity between the target and Hamiltonian-evolved states in the vacuum subspace in  $\text{Sec.~\ref{sec:main level 4}}$, we compute the logical fidelities between the two states in the GKP codespace. This is because logical lifetime is the primary figure of merit in quantum error correction schemes, and the subsystem decomposition captures the logical content spread across the codespace and error spaces, providing a lens to view how codestates are corrupted over time. Due to the numerical challenges of simulating ECD starting in fluxonium state superpositions in the presence of photon loss, we only simulate product states and average the resulting fidelities. Note that these fidelities do not include measurement-induced dephasing of the fluxonium due to photon loss in the resonator.

In Fig.~\ref{fig:modified ecd sequence}(a), we numerically evaluate logical fidelities for a target displacement $\beta = \sqrt{2\pi}$ as a function of gate time, obtaining fidelities close to $99\%$ under 600 ns. With $T_1/T_{\mathrm{ECD}} \gg 1$, this operation regime allows us to perform multiple rounds of error correction and further boost fidelities if needed. In Fig.~\ref{fig:modified ecd sequence}(b), we compare the GKP fidelities for a gate duration of 240 ns and vary the intermediate displacement $\alpha_0$, obtaining logical fidelities exceeding $99\%$. The strength of the improved ECD control sequence lies in the underlying theoretical modelling, which allows us to accurately track the resonator's position in phase space and compensate for deviations in its trajectory (see Fig.~\ref{fig:semiclassical trajectories} and Appendix~\ref{appendix:level2}). The model predicts the residual unconditional displacements and misrotations, which are then successfully removed by the improved ECD sequence (see Fig.~\ref{fig:average displacement}). We assume access to perfect calibration of displacements and extract the resultant conditional displacement by finding the coherent state amplitude of the resonator in all the above fidelity analyses. Specifically, we compute the logical fidelity of the density matrix $\rho$  evolved under ECD to a GKP state displaced by $\beta = \text{Tr($\rho\hat{a}$)}$  where $\ahat$ is the dressed resonator annihilation operator, isolating displacement errors from state deformations. Consequently, the performance of standard ECD and modified ECD protocols is expected to be identical in this scenario of perfect calibration of resonator position, as illustrated in Fig.~\ref{fig:modified ecd sequence}(a, b ).  

In the presence of resonator photon loss, we obtained fidelities up to $99-99.9\%$ in the GKP codespace $\left(\text{Fig.~\ref{fig:modified ecd sequence}(a, b)}\right)$. Since we observed weak fluxonium leakage in the device $\left(\text{see inset Fig.~\ref{fig:vacuum fidelities}(b)} \right)$, the degradation in fidelities in the GKP codespace potentially arises due to two main effects: photon loss and the resonator nonlinearities deforming the state, both bringing the state outside the codespace. At $\alpha_0 = 0$, the fidelity is limited by photon loss and resonator nonlinearities (Fig.~\ref{fig:modified ecd sequence}(b) inset). To isolate the contribution of photon loss against spurious nonlinearities, we simulate ECD dynamics in the same device setting $\kappa = 0$ and observe one order of improvement in fidelities in the absence of loss. This indicates that in the ideal fluxonium - lossy resonator device we numerically studied, photon loss limits the achievable ECD fidelities in the GKP code basis. 

The time required to attain a conditional displacement $\beta$ with an intermediate displacement $\alpha_0$ is given by $\approx \beta/\chi\alpha_0$~\cite{Eickbusch_2022}. In the presence of photon loss, the amplitude $\alpha$ of the coherent state $\ket{\alpha}$ decays like $\alpha \rightarrow \alpha e^{-\kappa T/2}$, slowing down the gate time. To remedy this, we further propose introducing a slow repumping drive on the resonator which continuously pumps the resonator back to the trajectory defined by $|\alpha| = \alpha_0$, speeding up ECD in the presence of photon loss $\left(\text{see Appendix~\ref{appendix:level5}}\right)$.
\section{\label{sec:main level 6}CONCLUSION}
We numerically implemented Echoed Conditional Displacement (ECD) gates in a resonator-fluxonium device and provided a comprehensive study of the resonator dynamics in the strongly driven regime. Our results demonstrate that it is necessary to go beyond lower-order perturbation theory when the resonator is subject to a strong drive due to the activation of higher-order terms through phase-space amplification. By describing the resonator in its displaced frame, we systematically account for the contributions of all-order terms in the dispersive expansion -- using a combination of semiclassical trajectories and master equation simulations. The displaced-frame treatment unveils non-trivial observations such as the saturation of the dispersive shift at large displacements, which leads to a non-monotonic dependence of the control speed on phase-space displacement. To evaluate the performance limits of ECD, we characterize the error mechanisms bounding the gate fidelity in a numerically simulated device featuring a lossy resonator and a dissipationless fluxonium control qubit. Our analysis confirms minimal leakage out of the fluxonium computational subspace post-ECD execution, and we identify resonator photon loss as the primary error mechanism limiting the achievable fidelities. We developed an improved ECD sequence that actively accounts and compensates for the effects of photon loss and spurious higher-order terms on resonator trajectories. The pulse sequence is analytically computed starting from the semiclassical resonator model we developed, providing a robust and analytic approach to improve the performance of ECD.

The effective, all-order perturbative treatment of the resonator carried out in the displaced frame presented in this work provides a framework to map the phase-space landscape of resonator dispersion, offering a pathway to identify optimal device operating points. Because the interplay between control speed and the phase-space landscape depends strongly on the choice of device parameters, our modeling establishes a systematic route to engineer the phase-space profile of ECD control speed during the device design stage.
\section{ACKNOWLEDGEMENTS}
The authors thank Benjamin D'Anjou for critical feedback on the manuscript. This work was financially supported by the Army Research Office under the grant W911NF2310045,  National Science and Engineering Council, and the Canada First Research Excellence Fund. A.A. acknowledges support from the FRQNT (Fonds de recherche du Québec – Nature et technologies) PBEEE $\left(\text{Bourses d’excellence pour étudiants(es) étrangers(ères)}\right)$ programme under Grant No. 2011518. S.R.J. and S.D.C. acknowledge support from the National Science Foundation Graduate Research Fellowship under Grant No. 1745302 and the CQE-LPS Doc Bedard Fellowship. Any opinion, findings, and conclusions or recommendations expressed in this material are those of the authors(s) and do not necessarily reflect the views of the National Science Foundation or LPS/ARO.
\appendix
\section{\label{appendix:level1}STATE IDENTIFICATION}
\begin{figure}[h]
    \centering
    \includegraphics[width=1\linewidth]{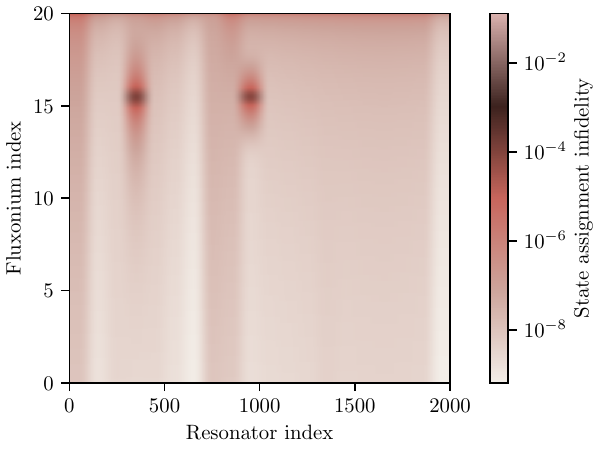}
    \caption{$\left(\text{a}\right)$ Overlap $\left(\text{approximate, see~\cite{Shillito_2022}}\right)$ of the bare product states $\ket{q}\otimes\ket{r}$ with dressed eigenstates identified with label $\ket{q, r}$.}
    \label{fig:state assignment}
\end{figure}

A robust state labeling procedure is crucial for extracting the right physical observables because all measurements occur within the dressed eigenbasis. Consequently, the dressed eigenstates of the coupled system must be identified with quantum numbers that map them to the bare states of individual subsystems. Within the dispersive regime, dressed eigenstates strongly resemble the bare product states, containing only minor contributions from other bare levels. To label the system's eigenbasis, we employ the state identification algorithm presented in Ref.~\cite{Shillito_2022}. In Fig.~\ref{fig:state assignment}, we show the state identification infidelity obtained using this technique, where the value corresponding to each point $(i, j)$ is approximately equal to one minus the overlap between the dressed eigenstate labeled as $\ket{n_q, n_r}$ and the bare product state $\ket{n_q}\otimes\ket{n_r}$. We obtain excellent state identification fidelities $>99.999\%$ in the computational manifold, enabling the construction of physically meaningful dressed observables.
\section{\label{appendix:level2}DISPLACEMENT TRANSFORMATION}
\begin{figure*}[tp]
    \centering
    \includegraphics[width=1\linewidth]{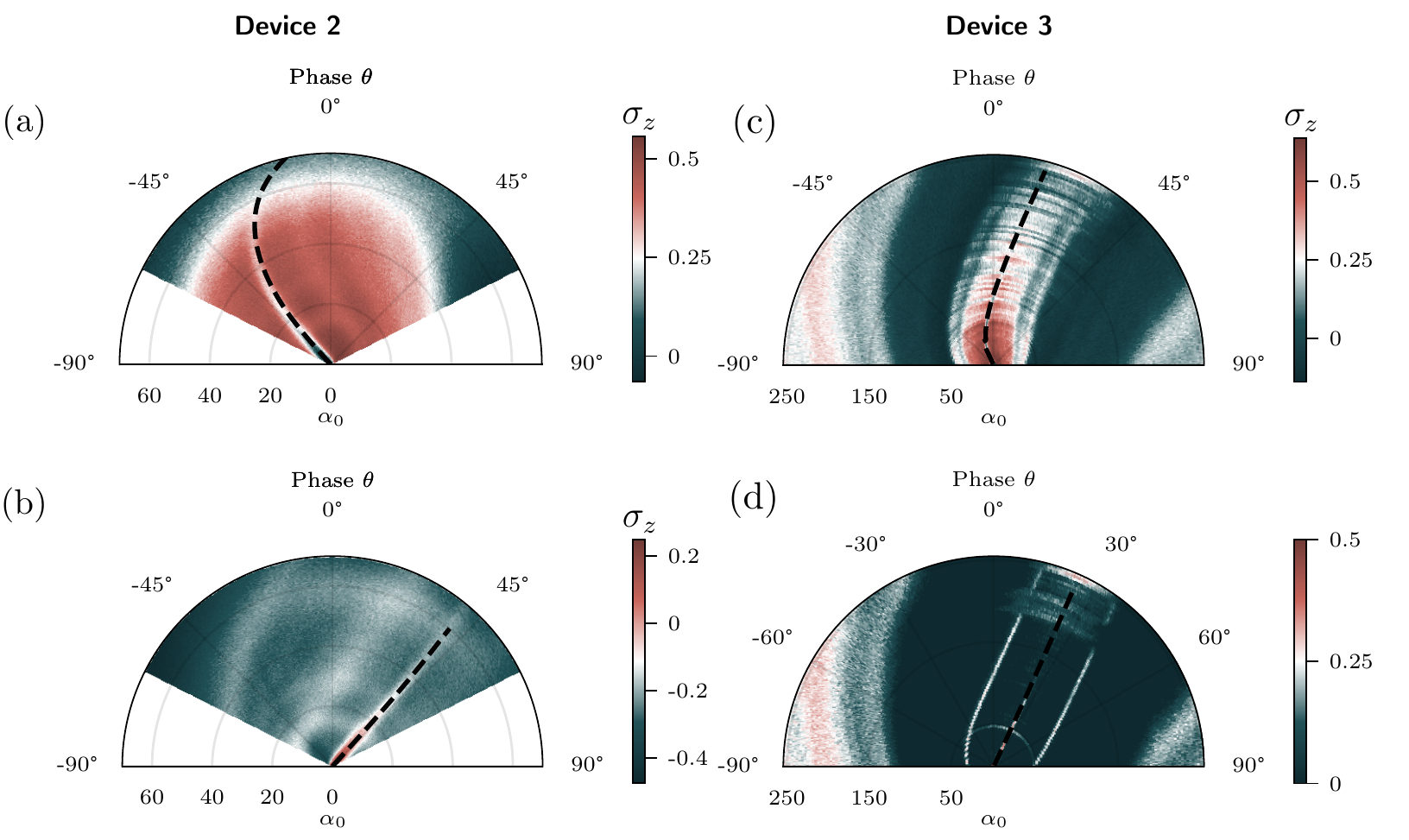}
    \caption{Numerical fits of \textit{out-and-back} (black dashed lines) match well with the experimentally measured spectrum (background). $\left(\text{a, b}\right)$ on the left are the results for Device 2 for fluxonium starting in $g$ (top) and $e$ (bottom). $\left(\text{c, d}\right)$ on the right are the results obtained for Device 3.}
    \label{fig:out_back_other_devices}
\end{figure*}
Let us consider the example of a dispersively coupled qubit-resonator system driven by a classical drive $\epsilon$ at a frequency $\omd$. The Hamiltonian in the co-rotating frame of the qubit and resonator can be written as
\begin{equation}
    \Hhat = \Delta\adag\ahat + \chi\adag\ahat\frac{\sz}{2} + \epsilon\left(\ahat + \adag\right),
    \label{dispersive_ecd_hamiltonian_2}
\end{equation}
with $\Delta = \left(\omr - \omd\right)$. We want to go to a frame where this Hamiltonian is diagonal and time-independent, which makes it straightforward to solve. This is the idea of the displacement transformation, where we move to a frame that follows the trajectory of the resonator in phase space at every instant of time such that the resonator appears to be static in this frame.  

We define the unitary $\Ud = e^{\alpha\left(t\right)\adag - \alpha\left(t\right)^*\ahat}$, which is a time-dependent displacement operator. Under $\hat{U}_d$, the Hamiltonian given in Eq.~\eqref{dispersive_ecd_hamiltonian_2} transforms as
\begin{equation}
    \Hhat \rightarrow \Ud\Hhat\Ud^{\dagger} - i\partial_t{\Ud}\Ud^{\dagger}.
    \label{unitary_transformation_rule}
\end{equation}
The Hamiltonian in this displaced frame takes the form
\begin{equation}
\begin{aligned}
    \Hhat =& \Delta\adag\ahat + \chi\adag\ahat\frac{\sz}{2} + \chi|\alpha|^2\frac{\sz}{2} + \\
    &\adag\left( i\partial_t\alpha + \Delta\alpha + \frac{\sz}{2}\chi\alpha +\epsilon\left(t\right)\right) + \text{h.c.}
    \label{displaced_ecd_dispersive_hamiltonian}
\end{aligned}
\end{equation}
Since the qubit does not change its state during a dispersive evolution, we can evaluate the Hamiltonian given in Eq.~\eqref{displaced_ecd_dispersive_hamiltonian} in the ground and excited qubit sectors separately. Choosing $\alpha$ given by the solution of 
\begin{equation}
     \partial_t\alpha = -i\Bigg[\left(\Delta + \chi\frac{\langle\sz\rangle}{2}\right)\alpha + \epsilon\left(t\right)\Bigg], 
     \label{alpha_ode}
\end{equation}
yields a diagonal Hamiltonian that can be easily solved.

Now we consider the case of a resonator capacitively coupled to a multilevel fluxonium described in Eq.~\eqref{rf_hamiltonian} of the main text. The full dispersive form of the Hamiltonian described by Eq.~\eqref{rf_hamiltonian} can be expressed as
\begin{equation}
\begin{aligned}
    \Hhat &= \Delta\adag\ahat - \sum_{n = 2}{K^{2n}\hat{a}^{\dagger n}\ahat^n}\\
    &- \sum_{n = 1, m = 0} \chi^{\left(2n + 2, m\right)}\hat{a}^{\dagger n}\ahat^n\ketbra{m}{m}.
\end{aligned}
\label{generic_dispersive_model}    
\end{equation}
Confining Eq.~\eqref{generic_dispersive_model} to the qubit manifold, we obtain
\begin{equation}
\begin{aligned}
\Hhat &= \Delta\adag\ahat - \sum_{n = 2}{K^{2n}\hat{a}^{\dagger n}\ahat^n} - \\
&\sum_{n = 1} \left(\chi^{\left(2n+2, g\right)}\ketbra{g}{g} + \chi^{\left(2n+2, e\right)}\ketbra{e}{e}\right)\hat{a}^{\dagger n}\ahat^n.
\end{aligned}
\label{generic_dispersive_model_qubit_manifold}
\end{equation}    
\noindent
Following the discussion above, under a drive $\epsilon$ and applying a displacement transformation ($\ahat\rightarrow \ahat + \alpha$) gives rise to new terms in the displaced frame. Higher-order terms that correspond to frequency shifts would get diluted, giving rise to amplified lower-order terms as follows:
\begin{equation}
    \begin{aligned}
        &K^{2n} \xrightarrow{\mathrm{lower-order \hspace{.1cm} amplification}} nK^{2n} |\alpha|^{2\left(n - 1\right)},\\
        &\chi^{\left(2n + 2, m\right)}  \xrightarrow{\mathrm{lower-order \hspace{.1cm} amplification}} n\chi^{\left(2n + 2, m\right)} |\alpha|^{2\left(n - 1\right)}.
    \label{phase_space_amplification}
    \end{aligned}
\end{equation}
We ignore number non-conserving terms arising in the displaced frame and only retain number-conserving terms that lead to the frequency dressing of the system. Using the relation $\sz = \ketbra{e}{e} - \ketbra{g}{g}$, we can rewrite Eq.~\eqref{generic_dispersive_model_qubit_manifold} as
\begin{equation}
\begin{aligned}
     &\hat{H}_{\mathrm{disp}} = \left[\deleff^\p\ \left( |\alpha|^2 \right) - \chi_{\mathrm{eff}}\left( |\alpha|^2\right)\frac{\sz}{2}\right]\adag\ahat, \\
\end{aligned}
    \label{effective_displaced_semiclassical_model}
\end{equation}
where
\begin{equation}
\begin{aligned}
    &\deleff^\p = \deleff - \frac{\chieff}{2},\\
    &\deleff\left(\modalpha\right) = \Delta - \sum_{n}^{}2nK^{\left(2n\right)}|\alpha|^{2\left(n - 1\right)},\\
    &\chieff(\modalpha) = \sum_{n}^{}2n|\alpha|^{2\left(n - 1\right)}\left(\chi^{\left(2n+2, e\right)} - \chi^{\left(2n+2, g\right)}\right).
    \label{effective_parameters}
\end{aligned}
\end{equation}
The parameters $\deleff\left(\modalpha\right)$ and $\chieff\left(\modalpha\right)$ capture the photon-number dependence of the resonator to all orders exactly, in principle. For brevity, we will relabel $\deleff^\p$ as $\deleff$ from now onwards. We numerically compute the effective parameters by a recursive extrapolation of photon-number dependent dispersion in the low-energy sector of the resonator. The phase-space trajectory of the resonator can then be obtained using
\begin{equation}
     \partial_t\alpha = -i\Bigg[\left(\deleff\left(\modalpha\right) - \chieff\left(\modalpha\right)\frac{\langle\sz\rangle}{2}\right)\alpha + \epsilon\left(t\right)\Bigg] .
     \label{resonator_trajectory}
\end{equation}

 For instance, we can retrieve the quartic dispersive Hamiltonian in Eq.~\eqref{dressed_rf_hamiltonian_dispersive} if we truncate the summation at $n = 2$. In this case, the effective Hamiltonian is given by
\begin{equation}
    \hat{H}_{\mathrm{disp}} = \left[\Delta  - 2K\modalpha - \frac{ \chi}{2}\sz\right]\adag\ahat, 
    \label{effective_dispersive_model_fourth_order}
\end{equation}
and the resonator state $\alpha$ is given by
\begin{equation}
     \partial_t\alpha = -i\Bigg[\left(\Delta - 2K\modalpha - \chi\frac{\langle\sz\rangle}{2}\right)\alpha + \epsilon\left(t\right)\Bigg] .
     \label{alpha_ode_dispersive_model_fourth_order}
\end{equation}

The simulation framework proceeds via an iterative, recursive loop summarized as follows:
\begin{enumerate}
    \item \textbf{Initial Estimation}: To determine the effective dispersion $\chi_{\mathrm{eff}}^{g(e)}$ at a specific photon number $\lvert \alpha_i \rvert^2$, we first generate an initial guess by extrapolating the Hamiltonian parameters obtained via numerical diagonalization in the low-photon manifold.
    
    \item \textbf{Displaced-Frame Simulation}: Using this guess, we solve the resonator trajectory in Eq.~\eqref{resonator_trajectory} and simulate the system Hamiltonian Eq.~\eqref{rf_hamiltonian} within the resulting displaced frame.
    
    \item \textbf{Recursive Update}: This Schrödinger equation simulation yields a refined value for the dispersion $\chi^{g(e)}_{\mathrm{eff}}\big(\lvert \alpha_i \rvert^2\big)$, which is then recursively fed back into the procedure.
\end{enumerate}

\begin{figure}[h!]
    \centering
    \includegraphics[width=1\linewidth]{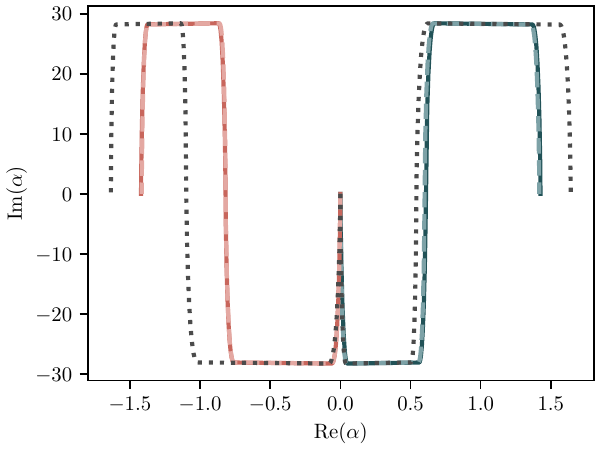}
    \caption{Phase-space trajectories of the resonator evolving under ECD are computed numerically. Solid lines denote trajectories obtained by solving the master equation with the qubit starting in $g \left(\text{teal}\right)$ and $e\left(\text{coral}\right)$, respectively. Dashed lighter lines overlaying the solid darker lines are obtained by solving the ODE in Eq.~\eqref{resonator_trajectory} given by our effective semiclassical model. Dotted gray lines represent the solutions to Eq.~\eqref{alpha_ode_dispersive_model_fourth_order} obtained using a fourth-order dispersive model.}
    \label{fig:semiclassical trajectories}
\end{figure}

Using the semiclassical equations of motion we derived above, we can numerically obtain the phase-space trajectories of the resonator evolving under an ECD sequence given by Eq.~\eqref{ecd_unitary}. In Fig.~\ref{fig:semiclassical trajectories}, we compare the semiclassical trajectories obtained from our effective model and the fourth-order dispersive form against the results obtained numerically by solving the time-dependent master equation, achieving remarkable agreement between our semiclassical theory and the master equation simulations. 

We numerically diagonalize the uncoupled fluxonium Hamiltonian in the Fock basis using 1500 states, retaining 20 eigenstates, and express the system Hamiltonian in Eq.~\eqref{rf_hamiltonian} in the eigenbasis of the bare fluxonium and bare resonator. Numerical Hilbert space sizes of the resonator and fluxonium were chosen after verifying the convergence of Hamiltonian parameters against the truncations of Hilbert space size. With the precise knowledge of the resonator trajectory, we can numerically simulate the driven resonator-fluxonium Hamiltonian expressed in Eq.~\eqref{rf_hamiltonian} in its time-dependent displaced frame. We continuously track the resonator trajectory so that the resonator remains nearly in its ground state at every instant, which allows us to efficiently simulate the strongly driven resonator-fluxonium system with a small numerical Hilbert space size. For Schrödinger equation simulations, we keep up to four resonator states in the displaced frame and 20 fluxonium states, yielding a total Hilbert space dimension of 80, and for master equation simulations we truncate to four resonator states in its displaced frame and eight fluxonium states, yielding a total Hilbert space dimension of 32. With such accurate frame tracking of the resonator, we numerically simulated the full driven resonator-fluxonium system described by Eq.\eqref{rf_hamiltonian} up to 50,000 photons, invoking no approximations.
\section{\label{appendix:level3}OUT-AND-BACK}
The \textit{out-and-back} protocol is an example of a phase-space enhancement technique that is used to measure resonator nonlinearities in the weak dispersive regime. The \textit{out-and-back} pulse sequence is illustrated in Fig.~\ref{fig:out and back}, where we start by initializing the fluxonium in either the ground or excited state, with the resonator in the vacuum state. A first drive $\epsilon$ displaces the resonator to a coherent state $\ket{\alpha}$ and, after a free evolution time $T$, the resonator state acquires a phase determined by the effective dispersion at that photon number given in Appendix~\ref{appendix:level2}, resulting in the state $\ket{\alpha e^{i\phi_{\alpha}}}$ with
\begin{equation}
    \phi_{\alpha} = \left(\deleff\left(\modalpha\right) - \chieff\left(\modalpha\right)\frac{\langle\sz\rangle}{2}\right)T.
    \label{out_and_back_phase}
\end{equation}
The resonator is then subjected to a second drive $-\epsilon\left(t\right)e^{i\theta}$, where $\theta$ is swept, and tracking the phase that returns the resonator to vacuum provides direct information about the resonator dispersion. However, if nonlinearities distort the coherent state into a highly deformed quantum state, no choice of phase can return the resonator to its initial vacuum state. Another scenario leading to this ``no-return'' condition occurs if the qubit transitions out of its computational basis by ionizing into higher-energy excited states, marking the breakdown of the dispersive regime.

\begin{figure}[h!]
    \centering
    \includegraphics[width=1\linewidth]{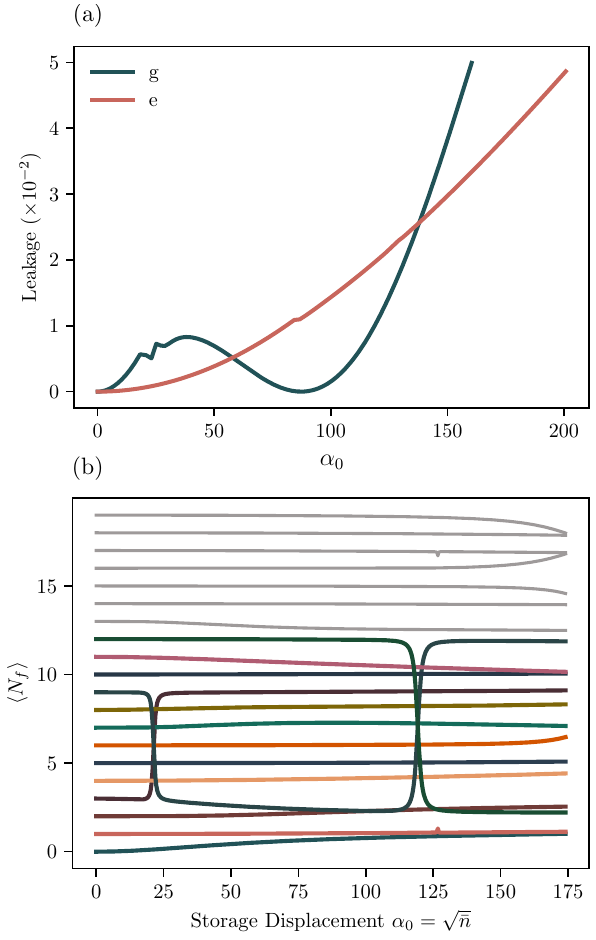}
    \caption{$\left(\text{a}\right)$ Fluxonium leakage population outside the computational states after \textit{out-and-back}, as a function of resonator displacement. $\left(\text{b}\right)$ Branch analysis of the simulated device shows no observable hybridization of higher states with the qubit branches $g$ and $e$. There is a very weak hybridization between $e$ and the $17$-th branch of fluxonium around $\alpha = 125$, which is unlikely to affect the ground state dynamics.}
    \label{fig:heating_leakage}
\end{figure}

In the numerically simulated \textit{out-and-back} experiment, signal loss occurs around \(\alpha \approx 170\) for the qubit starting in $\ket{g}$. To identify the exact mechanism causing the breakdown of the sequence, we numerically evaluate the post-protocol population leakage outside the computational manifold, for instances of correct drive return phases. Fig.~\ref{fig:heating_leakage}(a) shows the numerically extracted leakage populations computed from the reduced fluxonium density matrix of these instances, which remain below $5\%$. This confirms that the qubit remains predominantly in its initial state. Furthermore, branch analysis of the device shown in Fig.~\ref{fig:heating_leakage}(b) also indicates no sign of drive-induced state transitions ~\cite{Shillito_2022, Dumas_2024, Eickbusch_2022} around the ``no-return'' point. The signal degradation is likely caused by drive-enhanced, higher-order nonlinearities that deform the state and prevent its return to the vacuum.


\begin{table}[htbp]
    \centering
    \renewcommand{\arraystretch}{1.2} 
    \setlength{\tabcolsep}{8pt}       
    \caption{Device parameters used for the numerical simulations. Unless stated otherwise, all results in the main text and Appendix are obtained with the parameters of Device 1.}
    \label{Table_device_parameters}
    \begin{tabular}{|c|c|c|c|}
        \hline
        Parameter  & Device 1 & Device 2 & Device 3 \\
        \hline
        $\omega_r/2\pi$ & 4.146 GHz & 4.147 GHz & 4.138 GHz \\ 
        $E_C/2\pi$      & 856.8 MHz & 997.3 MHz & 787.8 MHz \\
        $E_L/2\pi$      & 301.8 MHz & 239.3 MHz & 564   MHz \\
        $E_J/2\pi$      & 2.7562 GHz& 2.797 GHz & 2.787 GHz \\
        $g/2\pi$        & 6.3 MHz   & 7.8  MHz  & 4.6   MHz \\
        \hline
    \end{tabular}
\end{table}

We numerically simulated the \textit{out-and-back} protocol across several devices used during the experiments in Ref.~\cite{Jha_2026} (parameters listed in Table~\ref{Table_device_parameters}). Our theoretical model consistently reproduced the experimental data with one adjustable parameter on the return phases in Device 2 and Device 3, demonstrating the generality and accuracy of our semiclassical formulation as illustrated in Fig.~\ref{fig:out_back_other_devices}. Introducing an adjustable parameter was expected due to the miscalibration of photon number when these experiments were performed on Devices 2 and 3. 
\section{\label{fidelities in vacuum subspace}EFFECT OF RESONATOR NONLINEARITIES ON VACUUM SUBSPACE}
\begin{figure}[t]
    \centering
    \includegraphics[width=1\linewidth]{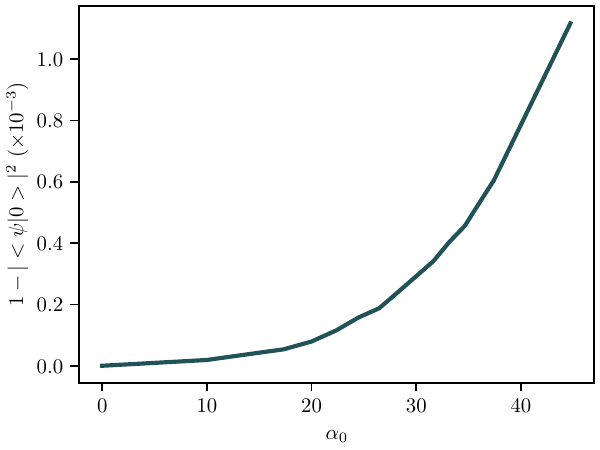}
    \captionsetup{width=1\linewidth}
    \caption{Numerically obtained vacuum overlap of the ECD return state, as a function of phase-space displacement $\alpha_0$ for a gate time of 240 ns.}
    \label{fig:overlap degradation}
\end{figure}

Numerical simulations within the vacuum subspace suggest that resonator nonlinearities fundamentally constrain achievable gate fidelities due to state deformation. To isolate this distortion from mistargeted displacements and state rotations, we compute the evolved state's coherent amplitude, $\beta = \langle\hat{a}\rangle$, using the dressed resonator annihilation operator $\ahat$. We then evaluate the vacuum overlap of the counter-displaced state, given by \(\langle0\vert{}\hat{D}^{\dagger}(\beta) \text{ECD}(\beta)\vert{}0\rangle\). If nonlinearities dictate the fidelity floor, this overlap should decrease because the accumulated deformation prevents a perfect return to the vacuum state. Infidelities in vacuum subspace (Fig.~\ref{fig:vacuum fidelities}(b)) are found to be on par with the degradation in state overlaps numerically obtained in Fig.~\ref{fig:overlap degradation}, confirming that state distortion due to the action of resonator nonlinearities is indeed the fidelity limiting factor in the vacuum subspace.
\section{\label{appendix:level4} ECD ACCOUNTING FOR NONLINEARITIES AND PHOTON LOSS}
\begin{figure}[h!]
    \centering
    \includegraphics[width=1\linewidth]{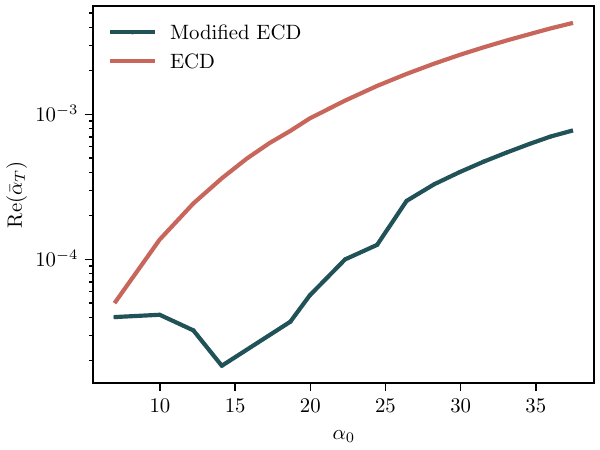}
    \captionsetup{width=1\linewidth}
    \caption{For a fixed gate time $T = 240$ ns, the average unconditional displacement $\bar{\alpha}\left(T\right)$ defined by Eq.~\eqref{unconditional_displacement_ecd} along the targeted conditional displacement axis $\text{Re$\left(\alpha\right)$}$ is numerically obtained as a function of phase-space displacement.}
    \label{fig:average displacement}
\end{figure}
Standard ECD drive amplitudes are derived assuming an ideal, lossless resonator with no self-Kerr. The action of higher-order nonlinearities and photon loss becomes relevant at high photon numbers. These terms results in state deformations and accumulation of  unconditional displacements and state rotations which leads to slightly different dynamics. This discrepancy between the idealized dispersive dynamics and the actual system evolution is typically resolved by opting for numerical optimization of the pulse parameters to construct ECD gates. Here, we introduce a modified ECD sequence that explicitly accounts for and compensates for the unconditional displacements and misrotations caused by resonator loss and higher-order nonlinearities, removing the need for complicated drive calibration procedures.

We consider a drive of the form
\begin{equation}
 \epsilon\left(t\right) = A_1\delta\left(t\right) + A_2\delta\left(t - \frac{T}{2}\right) + A_3\delta\left(t - T\right). 
\label{ecd_generic_pulse_sequence}
\end{equation}
The semiclassical trajectory for the resonator starting in vacuum and fluxonium in $\ket{g}$, evolving under the Hamiltonian expressed in Eq.~\eqref{rf_hamiltonian} with photons leaking outside the cavity at a rate $\kappa$, is given by
\begin{equation}
\partial_t \alpha_g = -i\Bigg[\deleff\argalpha + \frac{\chieff\argalpha}{2} - i\frac{\kappa}{2}\Bigg]\alpha_g -i\epsilon.
\label{resonator_phase_space_trajectory_ode}
\end{equation}
From now on, we drop the notation $\deleff\argalpha$ and $\chieff\argalpha$ and simply write $\deleff$ and $\chieff$ for brevity. The solution for the ODE in Eq.~\eqref{resonator_phase_space_trajectory_ode} is given by
\begin{equation}
\begin{aligned}
\alpha_g\left(t\right) =& \alpha_g(0)e^{-i\left(\deleff + \frac{\chieff}{2}\right)t}\\ - &\int_{0}^{t}\epsilon\left(\tau\right) e^{-i\left(\deleff+ \frac{\chieff}{2}\right)\left(t - \tau\right)}\,d\tau.
\end{aligned}
\label{resonator_phase_space_trajectory}
\end{equation}
For $0 \leq t < \frac{T}{2}$, we obtain
\begin{equation}
    \alpha_g\left(t\right) = -iA_1e^{-i \left(\deleff + \frac{\chieff}{2} - i\frac{\kappa}{2}\right)t}.
    \label{resonator_phase_space_trajectory_1}
\end{equation}
At $t = \frac{T}{2}$, the qubit state is flipped with a $\pi$ echo pulse. So $\sigma_z \rightarrow -\sigma_z$, or equivalently $\chi \rightarrow -\chi$, for $t>\frac{T}{2}$.
For $\frac{T}{2} < t < T$, we obtain
\begin{equation}
\begin{aligned}
\alpha_g\left(t\right) =& \alpha_{g}\left(\frac{T}{2}\right) e^{-i\left(\deleff - \frac{\chieff}{2} - i\frac{\kappa}{2}\right)(t - \frac{T}{2})} \\ 
&-iA_2 e^{-i\left(\deleff - \frac{\chieff}{2} - i\frac{\kappa}{2}\right)\left(t - \frac{T}{2}\right)}.
\label{resonator_phase_space_trajectory_2}
\end{aligned}
\end{equation}
The resonator state at $t = \frac{T}{2}$ is given by
\begin{equation}
    \alpha_g\left(\frac{T}{2}\right) = -i\Big[A_1e^{-i\left(\deleff + \frac{\chieff}{2} - i\frac{\kappa}{2}\right)\frac{T}{2}} + A_2\Big].
    \label{resonator_phase_space_trajectory_3}
\end{equation}
For $t > T$, we obtain
\begin{equation}
\begin{aligned}
     \alpha_g\left(t\right) =&\, \alpha_{g}\left(\frac{T}{2}\right) e^{-i\left(\deleff - \frac{\chieff}{2} - i\frac{\kappa}{2}\right)\left(t - \frac{T}{2}\right)} \\
&-iA_2 e^{-i\left(\deleff - \frac{\chieff}{2} - i\frac{\kappa}{2})(t - \frac{T}{2}\right)} \\
&- iA_3 e^{-i\left(\deleff - \frac{\chieff}{2} - i\frac{\kappa}{2}\right)\left(t - T\right)}.
\end{aligned}
\label{resonator_phase_space_trajectory_4}
\end{equation}
Using a similar derivation in the case where the fluxonium qubit is initially in the excited state, the final state of the resonator is given by
\begin{equation}
\begin{aligned}
    \alpha_{g\left(e\right)}\left(T\right) =& -i\Big[A_1e^{-i\left(\deleff - i\frac{\kappa}{2}\right)T} + \\&A_2e^{-i\left(\deleff \mp \frac{\chieff}{2} - i\frac{\kappa}{2}\right)\frac{T}{2}}+ A_3\Big].
\end{aligned}  
\label{resonator_phase_space_trajectory_e}
\end{equation}
We need to solve for the pulse amplitudes $A_i = |A_i|e^{i\theta_i}$ for $i = 1, 2, 3$, where the six unknowns are to be found using Eq.~\eqref{resonator_phase_space_trajectory_3} and Eq.~\eqref{resonator_phase_space_trajectory_e} for fluxonium states $g$ and $e$. We now impose conditions such that the above equations of motion generate a purely conditional displacement interaction. We first set the amplitude of $A_1 = \alpha_0$, fixing the intermediate phase-space displacement of the resonator ($\text{if $\kappa$ = 0, then $|\alpha\left(t\right)| = \alpha_0$ at all times}$). We want the unconditional displacement to be zero at the final time $t = T$.
\begin{figure}[t]
    \centering
    \includegraphics[width=1\linewidth]{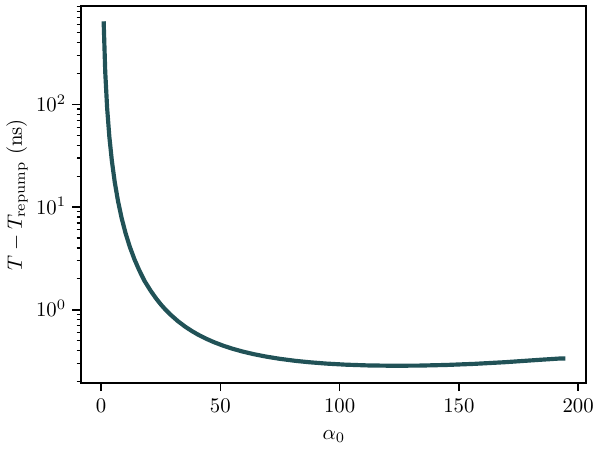}
    \captionsetup{width=1\linewidth}
    \caption{Numerically computed gate time difference $T - T_{\mathrm{repump}}$ versus the intermediate resonator displacement for a target $\beta = \sqrt{2\pi}$.}
    \label{fig:repumping ecd}
\end{figure}
\begin{equation}
\bar{\alpha}\left(T\right) = \frac{\alpha_g\left(T\right) + \alpha_e\left(T\right)}{2} = 0,  
\label{unconditional_displacement_ecd}
\end{equation}
which gives us the expression
\begin{equation}
A_1 e^{-i\left(\deleff - i\frac{\kappa}{2}\right)T} + A_2 e^{-i\left(\deleff - i\frac{\kappa}{2}\right)\frac{T}{2}}\cos{\left(\frac{\chieff T}{4}\right)} + A_3 = 0.
\label{unconditional_displacement_ecd_b}
\end{equation}
By imposing phase-matching conditions, we relate the complex phases between the terms in Eq.~\eqref{unconditional_displacement_ecd} as
\begin{equation}
    \theta_1 - \deleff T = \theta_2 - \frac{\deleff T}{2} = \theta_3.
    \label{phase_matching} 
\end{equation}
The conditional displacement $\beta^\p = \alpha_g(T) - \alpha_e(T)$ is given by
\begin{equation}
    \beta^\p = 2A_2 e^{-i\left(\deleff - i\frac{\kappa}{2}\right)\frac{T}{2}}\sin{\left(\frac{\chieff T}{4}\right)} = \beta e^{i\deleff T}.
    \label{ecd_generic_pulse_sequence0}
\end{equation}
The resonator state acquires a net rotation due to the effective detuning term $\deleff$, which rotates the gate by an angle $\deleff T$. To compensate for this effect, we deliberately target a conditional displacement underrotated by $\deleff T$, $\beta^\p = \beta e^{i\deleff T}$. This will yield the desired conditional displacement $\beta$ upon an overall frame rotation by $\deleff^\p T$, canceling the net rotation successfully.

Imposing phase-matching in Eq.~\eqref{ecd_generic_pulse_sequence0} gives us 
\begin{equation}
    \theta_2 - \frac{\deleff T}{2} = \theta_{\beta} + \deleff T.
    \label{ecd_generic_pulse_sequence1}
\end{equation}
We have found a relation that links the phase of the final conditional displacement $\beta$ to the drive phases. Now we proceed to finding the amplitudes $|A_i|$.

Due to photon loss acting on the resonator, $|\alpha(t)|$ does not remain on the circle of radius $|\alpha_0|$. At $t = \frac{T}{2}$, we want the state to be repumped back to $\alpha_0$, which imposes the condition $|\alpha_{g(e)}(T/2)| = |\alpha_0|$:
\begin{equation}
|\alpha_0| =  |-i\alpha_0 e^{-i \left(\deleff + \frac{\chieff}{2} - i\frac{\kappa}{2}\right)\frac{T}{2}} - iA_2|.
\label{ecd_generic_pulse_sequence2}
\end{equation}
This is a quadratic equation with two solutions. One of the solutions gives $A_2 = 0$ in the limit $\kappa \rightarrow 0$, and excluding this gives us the solution for $|A_2|$ given by
\begin{equation}
    \begin{aligned}
    |A_2| =& -|\alpha_0| \Bigg [\cos{\left(\frac{\chieff T}{4}\right)}  e^{-\frac{\kappa T}{4}} \\& + \sqrt{1 - \sin^2\left({\frac{\chieff T}{4}}\right) e^{\frac{-\kappa T}{2}}} \Bigg] \nonumber.
    \label{ecd_generic_pulse_sequence3}
    \end{aligned}
\end{equation}
With $A_1$ and $A_2$ known, $A_3$ is determined from Eq.~\eqref{unconditional_displacement_ecd_b}.

\begin{figure*}[tbp]
    \centering
    \includegraphics[width=1\linewidth]{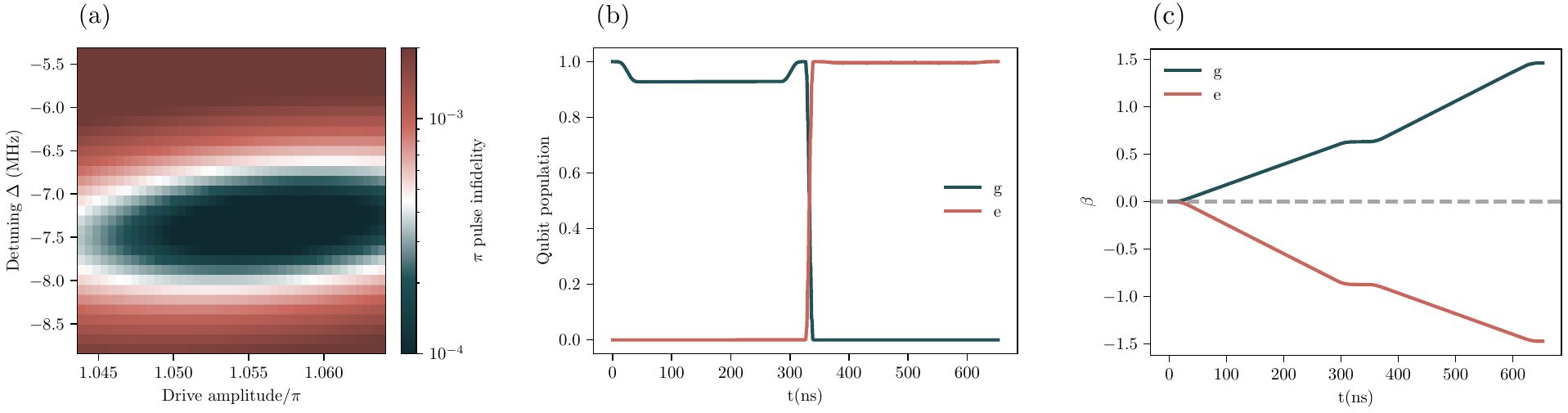}
    \caption{Simulating an $X$ gate with a drive on fluxonium. $\left(\text{a}\right)$  $\text{2D}$ frequency-amplitude scan of fluxonium drive  with the qubit initialized in $\ket{g}$. The colormap represents $1 - P_e$, where $P_e$ is the population in $\ket{e}$ $\left(\text{b}\right)$ Numerically obtained population transfer with $\pi$ pulse during ECD $\left(\text{c}\right)$ Conditional displacement accumulated during the pi pulsed ECD (with an intermediate displacement $\alpha_0 = 30$).} 
    \label{fig:pi pulse scan}
\end{figure*}

We have fully determined the drive coefficients required to generate an ECD gate with a conditional displacement of \(\beta e^{i\Delta_{\mathrm{eff}} T}\). More generally, to generate an arbitrary conditional displacement $\vert{}\beta\vert{}e^{i\theta_{\beta}}$ with an intermediate displacement amplitude of $\vert{}\alpha_0\vert{}$, the required drive amplitudes $A_{i}$ are given by

\begin{subequations}
\begin{align}
    A_1 &= |\alpha_0| e^{i \left(\theta_{\beta} + 2\deleff T\right)}, \\
    A_2 &= -|\alpha_0| \Bigg [\cos{\left(\frac{\chieff T}{4}\right)}  e^{-\frac{\kappa T}{4}} + \nonumber \\ 
    &\quad \sqrt{1 - \sin^2\left({\frac{\chieff T}{4}}\right) e^{\frac{-\kappa T}{2}}}\Bigg] e^{i \left(\theta_{\beta} + \frac{3}{2}\deleff T\right)}, \\ 
    A_3 &= -\left\{A_1 e^{-i\left(\deleff  - i\frac{\kappa}{2}\right)T}\right. \nonumber \\
    &\quad \left.+ A_2 e^{-i(\deleff - i\frac{\kappa}{2})\frac{T}{2}}\cos{\left(\frac{\chieff T}{4}\right)} \right\}.
\end{align}
\end{subequations}

In Fig.~\ref{fig:average displacement}, we compute the average unconditional displacements obtained with standard ECD against our modified ECD sequence and observe almost one order of suppression of residual displacements along the conditional displacement axis with our modified ECD protocol. The remaining unconditional displacements are due to the finite time of the pulses simulated.
\section{\label{appendix:level5}SPEEDING UP ECD WITH CONTINUOUS REPUMPING}
The strength of a conditional displacement interaction of intermediate displacement $\alpha _{0}$ is
\begin{equation}
    g_{\mathrm{CD}} = \chieff\left(|\alpha_0|^2\right)|\alpha_0|,
    \label{ecd_rate}
\end{equation}
and the total time required to achieve a target conditional displacement $\beta$ is given by
\begin{equation}
    T_{\mathrm{ECD}, \kappa = 0} {} = {} \frac{\beta}{g_{\mathrm{CD}}} = \frac{\beta}{\chieff\left(|\alpha_0|^2\right)|\alpha_0|}.
    \label{t_ecd}
\end{equation}
In the presence of photon loss $\kappa$, $\alpha \rightarrow \alpha e^{-\kappa T/2}$ and the resonator trajectory does not stay in the circle defined by $|\alpha| =|\alpha_{0}|$, slowing down the gate time. To counteract this decay, a continuous repumping drive can be applied to compensate for the dissipation, which forces the resonator to remain on the trajectory given by $\alpha_0$ throughout the gate execution, thereby recovering $T_{\mathrm{repump}} = T_{\mathrm{ECD, \kappa = 0}}$ in the presence of resonator loss. The analytical form of the drive that accomplishes the repumping can be derived from the semiclassical equation of motion for the resonator. Our objective is to design a drive $\epsilon_{\mathrm{rp}}$ that constrains the trajectory of the lossy resonator to match the ideal trajectory traced by a lossless system. Under a loss rate $\kappa$, the phase-space equation of motion for the resonator is expressed as,
\begin{equation}
\begin{aligned}
     \partial_t\alpha_{\mathrm{lossy}} &= -i\Bigg[\left(\deleff - \chieff\frac{\langle\sz\rangle}{2} - i\frac{\kappa}{2}\right)\alpha_{\mathrm{lossy}} + \epsilon_{\mathrm{rp}}\Bigg],
     \label{resonator_phase_space_ode_lossy}
\end{aligned}
\end{equation}
and this has to be equal to the resonator's path when there is no loss,
\begin{equation}
    \partial_t\alpha_{\mathrm{lossless}} = -i\Bigg[\left(\deleff - \chieff\frac{\langle\sz\rangle}{2}\right)\alpha_{\mathrm{lossless}} + \epsilon\Bigg],
    \label{resonator_phase_space_ode_lossless}
\end{equation}
where $\epsilon$ is the ECD drive given by Eq.~\eqref{ecd_generic_pulse_sequence}. From the condition $\alpha_{\mathrm{lossy}} = \alpha_{\mathrm{lossless}}$, we obtain the form of the repumping drive $\epsilon_{\mathrm{rp}}$:
\begin{equation}
\epsilon_{\mathrm{rp}} = \epsilon + i\frac{\kappa}{2}\alpha_{\mathrm{lossless}}.
\label{repumping_condition}
\end{equation}
We numerically compute the gate time $T$ required to implement a conditional displacement $\beta = \sqrt{2\pi}$ in the presence of photon loss using Eq.~\eqref{ecd_generic_pulse_sequence0}. Fig.~\ref{fig:repumping ecd} illustrates the gate time reduction achieved by the repumping protocol, quantified by the difference $T - T_{\mathrm{repump}}$. In the experimentally relevant regime of $\alpha \in [10, 20]$, the repumping scheme shortens the gate execution time by up to 150 ns.
\section{\label{appendix:level6}IMPLEMENTING THE $X$ GATE WITH A PI PULSE}
A $\pi$ rotation on the qubit Bloch sphere is achieved by applying a drive of the form $2\Omega(t)\cos{\omega_q t}$ through the fluxonium charge line. This requires that
\begin{equation}
    \begin{aligned}
        \bra{e, 0}\hat{N}_f\ket{g, 0}\int_{t_0}^{t_{\mathrm{final}}} \Omega(t) \,dt = \frac{\pi}{2}.
    \end{aligned}
    \label{fluxonium_pi_pulse_condition}
\end{equation}
For the qubit parked at the half-flux sweet spot, we use cosine flat-top pulses with a flat-top length of 14 ns and 4 ns of cosine rise and fall times. The resulting population transfer fidelity within the qubit subspace of the fluxonium exceeds $99.99\%$, as shown in Fig.~\ref{fig:pi pulse scan}(a). In Fig.~\ref{fig:pi pulse scan}(b, c) we illustrate results of a pi-pulsed ECD targeting an intermediate displacement $\alpha_0 = 20$, numerically obtaining $99.97\%$ fidelity in the vacuum subspace.
\FloatBarrier 
%

\end{document}